\documentclass{aastex631}

\shorttitle{Structure of the M31 Bulge}
\shortauthors{Leahy Craiciu \& Postma}
\graphicspath{{./}{figures/}}

\begin{document}

\title{The Complex Structure of the Bulge of M31}

\author{Denis Leahy }
\affiliation{Dept. Physics and Astronomy, University of Calgary, Calgary, AB, Canada;
}
\author{Traian Craiciu}
\affiliation{Dept. Physics and Astronomy, University of Calgary, Calgary, AB, Canada;
}
\author{Joseph Postma}
\affiliation{Dept. Physics and Astronomy, University of Calgary, Calgary, AB, Canada;
}

\begin{abstract}
The bulge of M31 is of interest regarding the nature of galactic bulges and how their structure relates to bulge formation mechanisms and their subsequent evolution.
With the UVIT instrument on AstroSat, we have observed the bulge of M31 in five far ultraviolet (FUV) and near ultraviolet (NUV) filters at 1" spatial resolution. 
Models for the luminosity distribution of the bulge are constructed using the UVIT data and the galaxy image-fitting algorithm GALFIT.  
We fit the bulge without the nuclear region with a Sersic function for the five images and find Sersic indices ($\simeq2.1$ to 2.5) similar to previous studies but smaller $R_e$ values ($\simeq0.5$ to 0.6 kpc).
When fitting the images including the nuclear region, a multi-component model is used. 
We use an 8-component model for the FUV 148nm image, which has the highest sensitivity. 
The other images (169 to 279 nm) are fit with 4-component models.
The dust lanes in the bulge region are recovered in the residual images, which have subtraction of the bright bulge light using the multi-component models. 
The dust lanes show that M31's nuclear spiral is visible in absorption at NUV and FUV wavelengths.
The bulge images show boxy contours in all five UVIT wavebands, which is confirmed by fitting using GALFIT. 
The Sersic indices of $\sim$2.1-2.5 are intermediate between the expected values for a classical bulge and for a pseudobulge. The boxiness of the bulge provides further evidence that M31's bulge has contributions from a classical bulge and a pseudobulge.
\end{abstract}

\keywords{Andromeda Galaxy (39); Galaxy bulges(578); Ultraviolet astronomy(1736); Galaxy structure(622); Galaxy formation(595)}

\section{Introduction} \label{sec:intro}

Galaxies in the local universe have clear morphological structures including bulges, disks and galactic halos\footnote{Galactic halos refers to the extended outer parts of galaxies, which are distinct from the dark matter halos that are well studied in cosmological simulations and which provide the gravity wells for early galaxy formation.}. 
In the widely accepted LCDM 
cosmology, galaxies, and their structures, are formed through initial density perturbations in dark matter halos followed by hierarchical build-up (accretion) of gas and various fragments of different sizes.
In the early Universe, galactic evolution was governed by rapid and violent mergers, dissipative collapse and hierarchical clustering. 
As the Universe expanded and merger events became less common, rapid evolutionary processes gave way to slow and persistent processes. 
For galactic bulges, the review of \citet{2004ARA&A..42..603K} (hereafter KK04) summarizes bulge structure, morphology and formation/evolutionary processes.
KK04 posit that in the late Universe, galactic evolution manifests primarily through internal secular processes. 
This occurs through the slow and steady interaction between the various internal galactic components such as the nucleus, bulge, bar and spiral arms. 

Simulations are used  to study the formation of disks and spheroids (bulges and halos) in the currently accepted LCDM cosmology (e.g. the simulations of \citealt{2011MNRAS.417..154S} for Milky Way mass haloes). 
From such simulations  the general picture is that bulges form early in the history of a galaxy and that disks are relatively young (e.g. \citealt{2016ASSL..418..317B, 2016MNRAS.459.4109T}).
Other studies focus on observational characteristics of galaxies and their structure. 
E.g., \cite{2011ApJ...742...96W} examine the dependence of galaxy structure on galaxy position in the star formation rate (SF rate or SFR) versus mass diagram\footnote{This diagram separates galaxies into two main categories: star forming galaxies, referred to as the SF main sequence and passive (non-star forming) galaxies, referred to as the red clump. 
Observations show that the difference in structure between the two types of galaxies is already in place by redshift $\sim$2.5.}. 
Other relevant results from \cite{2011ApJ...742...96W} are: the  Sersic index is $n\sim 1$ to 1.5 for SF main sequence galaxies and $n\gtrsim3$ for red clump galaxies. 
The observational study by \cite{2021A&A...656A.133Q} examines galactic structures and their relation to molecular gas and star formation in 
nearby galaxies.

Two categories of galactic bulges emerge from the dichotomy between galactic evolution in the early and late Universe (KK04). 
The “classical” bulge is interpreted within the framework of the Hubble-Sandage-de-Vaucouleurs classification scheme to represent an “elliptical living in the middle of a disk”, probably formed through a merger event. 
In contrast, if a bulge's evolution is dominated by internal secular processes it becomes a “pseudobulge”. 
While KK04 recognized difficulties in classifying intermediate cases, the basic interpretation is: 
if a bulge has elliptical shape it is a classical bulge, while a bulge with disk-like properties is a pseudobulge. 
In this manner, M31's bulge has been classified as a classical bulge, along with other galaxies such as M81 and the Sombrero galaxy (NGC 4594), which show bulges that are rounder than their associated disks. 

Evidence in support of the classical bulge formation theory has existed since the early simulations of \citet{1977egsp.conf..401T}. 
Evolutionary formation simulations by \citet{2003ApJ...597..893N} and \citet{2005A&A...437...69B} suggest that mergers of same-size disk galaxies have as a likely outcome a larger elliptical galaxy, while a greater mass ratio for the colliding disk galaxies results in spiral galaxy with a large elliptical bulge and with Sersic index $n<2$.
This supports the theory of the bulge being an elliptical galaxy residing within a larger spiral structure. 
A number of major merger simulations have found that the bulge of the resulting galaxy is a pseudo-bulge, with significant rotation, e.g., for the merger of two gas-rich pure disk galaxies \citep{2012MNRAS.424.1232K}. 
The major merger study by \citet{2016ApJ...821...90A} finds that 
most stars in the disks of merging galaxies form a classical bulge, with the remaining stars and stars born during the merger forming a thick disk and bar. After the merging phase a disky pseudobulge and the thin disk grow. 
The thin disk has the youngest stars, the pseudobulge has a wide range of ages, and the classical bulge has the oldest stars. 

It has been proposed that bulges can also take shape from instabilities in the early disk with large clumps and star clusters collapsing inward towards the center due to dynamical friction.  
\citet{1999ApJ...514...77N} describes galactic bulge formation by the inward transport of disk matter in the early history of a galaxy. 
In a chemo-dynamical simulation of galaxy formation, \citet{2004A&A...413..547I} found that when the cold gas in the nascent galaxy cools efficiently, massive clumps form that spiral inward to form a central bulge.
In contrast, slow cooling of the initial cold clouds leads to star formation in a more quiescent manner, with instabilities setting in later and resulting in a smaller bulge with a prominent stellar bar. 
The idea that clumps in the galaxy interact with the disk to provide fuel for both spiral and bulge structure is supported by 
study of \citet{2005A&A...437...69B}.  

The problem with bulge classification stems from the fact that internal secular evolution leads to an accumulation of dense components in the centers of disk galaxies that look similar to bulges built from classical mergers. 
The pseudobulges of the former were made slowly from disk gas, or largely influenced by the dynamics of the internal galactic components \citep{2013PASA...30...27M}. 
KK04 prescribe several key factors that can be used to identify a pseudobulge: 
it has a flattened shape, similar to its corresponding disk; it contains a nuclear bar (in relatively face-on galaxies); 
it is box-shaped (in relatively edge-on galaxies);  it has a Sersic index between 1 and 2; 
and it is more rotation dominated than classical bulges. 
It is apparent that a simple morphological assessment will not suffice and one must implement a quantitative fit to surface photometry to differentiate between classical and pseudo-bulges \citep{2013PASA...30...27M}.

For the detailed study of galactic structure, dynamics and evolution, there is no better astrophysical laboratory than our closest neighbouring spiral galaxy, the Andromeda Galaxy (M31), at a distance of 785$\pm$25 kpc \citep{2005MNRAS.356..979M}.
Our external view of M31 and its high Galactic latitude (thus low Galactic extinction) allow studies of the M31 bulge with a clearer view than comparative studies of our own Galactic bulge. 

There are several lines of evidence suggesting that internal secular evolutionary processes have taken place throughout M31’s history and have had a significant impact on its structure and dynamics. 
\citet{2006ApJ...638L..87G} find an offset star-forming ring in the bulge region outside of the nuclear spiral that is interpreted as a strong indicator of internal interactions between the structural components of M31. 
A mid-infrared survey of the morphology and surface brightness profiles of M31’s components by \citet{2006ApJ...650L..45B} shows the complex interactions between the nuclear region, bulge and disk.  
In HI mapping by \citet{2009ApJ...705.1395C}, several ring-like and spiral-like structures are observed with additional HI structures discovered in the outskirts of the disk, all thought to be the result of complex internal interactions. 

On the other hand, a history of external hierarchical formation for M31 is supported by several studies. 
Evidence for past interactions between M31 and its nearby dwarf companions M32 and NGC205 was given by \citet{2001Natur.412...49I}. 
The presence of many substructures in the halo are taken as evidence of numerous ($\sim$16) accretion events involving dwarf galaxies  \citep{2010ApJ...708.1168T}. 
A study of the rotational kinematics and metallicities of the M31 globular cluster system  \citep{2010MNRAS.401L..58B} concluded that an ancient major merger event is highly probable. 
 \cite{2014MNRAS.442.2929V} analyze the kinematics of the outer halo globular cluster system of M31, 
 concluding that a significant fraction of the globulars in the outer halo were accreted with their parent dwarf galaxies.
 Simulations by \cite{2018MNRAS.475.2754H} conclude that a major merger $\sim$2-3 Gyr ago explains the long-lived 10kpc star forming ring in M31, the absence of a remnant for the Giant Stream, and the spatial distribution of other major structures in M31's halo. 
\cite{2018NatAs...2..737D} argue that many of M31's unusual features (disk heating, tidal structures, massive and metal-rich halo) can be explained by a dominant merger $\sim$2 Gyr ago, with M32 the remnant core of the merger progenitor.   
 \cite{2018ApJ...868...55M} analyze the structure of M31's halo 
 to find that the 13 most distinctive substructures were produced by at least 5 different accretion events, all in the last 3 or 4 Gyr, and that a few of the large structures may have been produced by a single event. 

Thus there is plentiful evidence for the significance of mergers and of secular evolution over M31's history. 
Analyses of structural parameters for M31's bulge, relevant to the classical vs. pseudo-bulge question, have been carried out using observations in infrared and optical.
\cite{2008MNRAS.389.1911S} present an updated mass model for M31  
and find adiabatic collapse of the halo is favoured for M31, which favours secular (rather than merger driven) bulge formation, as the gradual accumulation of central mass increases the likelihood that adiabatic collapse will operate.
The analysis of \cite{2010A&A...509A..61S} 
finds the bulge is mainly old and of solar metallicity ($\gtrsim$12 Gyr) with a younger component in the inner few arcsec with metallicity $\sim$3 times solar.
The existence of mixed stellar populations in the bulge is confirmed by subsequent studies (e.g. \citealt{2018MNRAS.478.5379D,2022AJ....163..138L}). 
\cite{2011ApJ...739...20C}  
obtain structural parameters for its Sersic bulge (Sersic index ($n\simeq2$), effective radius $R_e\simeq1$ kpc), exponential disk and power-law-with-core halo. 
Based on ratio of bulge-to-disk scale lengths and on the Sersic index, they conclude that M31 likely evolved through early “classical” merging with structural readjustments caused by secular evolution and satellite accretion. 
\cite{2013ApJ...779..103D} find 
$n\simeq2$, a smaller bulge ($R_e = 0.77 \pm 0.03$ kpc) and varying luminosity functions with radius.
\cite{2018A&A...611A..38O} 
show that kinematics data support a bar in M31, with non-triaxial streaming motions. 
The presence of a bar is confirmed by \cite {2022ApJ...933..233F} in the central 4.6 by 2.3 kpc region of M31.
\cite{2018A&A...618A.156S} 
find that the bar and bulge are similar in age and [$\alpha$/Fe] maps, but the bar stands out in metallicity, approximately solar, in contrast to the metal-rich bulge.  
\cite{2018MNRAS.481.3210B} construct models for the M31 bulge including classical bulge and box/peanut bulge 
and find that both components are required to fit the data. 

In this study we analyze the structure of M31's bulge at ultraviolet wavelengths. 
The data is from the M31 UVIT survey \citep{2020ApJS..247...47L}, which was carried out  in  far ultraviolet (FUV) and near ultraviolet (NUV) by the UVIT instrument \citep{2017AJ....154..128T}  on AstroSat \citep{2014SPIE.9144E..1SS}.
Some previous results on UVIT FUV and NUV studies of the M31 bulge include: 
detection of young hot stars in the bulge \citep{2018AJ....156..269L}; 
analysis of the radial surface brightness distribution of the bulge \citep{2021IJAA...11..151L}; 
detection and characterization of FUV variable sources in the bulge \citep{2021AJ....161..215L} by comparison of two epochs of observation; 
analysis of the FUV to far infrared spectral energy distribution of the M31 bulge to study the star formation history of M31 \citep{2022AJ....163..138L}; and a SED analysis of the UVIT and PHAT photometry of clusters in the northern disk of M31 \citep{2022AJ....164..183L}.

The structure of the bulge of M31 is analyzed in detail here using the 2 dimensional surface brightness modelling software GALFIT \citep{2002AJ....124..266P}.
In Section~\ref{sec:obs} we briefly describe the FUV and NUV  images of the bulge of M31.  
Section~\ref{sec:galfit} describes GALFIT briefly and the process of modeling the surface brightness distribution of the M31 bulge.   
Section~\ref{sec:bulgemodel} describes the process and the results of modelling the outer bulge omitting the  the nuclear region's light distribution, and Section~\ref{sec:nucmodel} presents the resulting models incorporating the nuclear region. 
Position offsets between the centers of the components and of the nucleus, and bulge asymmetry are 
discussed in Section~\ref{sec:offset}. 
GALFIT’s performance is reviewed in Section~\ref{sec:reliability}.
Comparison with previous measurements of the structure of M31's bulge at other wavelengths is given 
in Section~\ref{sec:compare}, and the phenomenological nature of our multicomponent fits is discussed 
in Section~\ref{sec:interp}.
The classical vs. pseudobulge nature of M31's bulge is discussed in Section~\ref{sec:bulgetype} including 
Sersic index and boxiness indicators. 
We reveal the nuclear spiral in FUV and NUV wavebands using the model-subtracted residual images in Section~\ref{sec:spiral}, 
and conclude with a summary in Section~\ref{sec:conc}. 

\section{Observations and Bulge Modeling} \label{sec:methods}

\subsection{Observations}\label{sec:obs}

The AstroSat observatory and its science capabilities are described by \citet{2014SPIE.9144E..1SS}.
The UVIT instrument, its filters and its calibration are described by \citet{2017AJ....154..128T}.
UVIT has 1" spatial resolution, and a cicular field of view with diameter $\sim28$ arcmin. 
Observations were obtained in FUV, with F148W and F172M filters, and in NUV, with N219M and N279N filters
   as part of the M31 UVIT survey \citep{2020ApJS..247...47L}.
 The survey covered the sky area of M31 with 19 different fields, with planned exposures in FUV and NUV filters for each field. 
 The filters used were F148W, F169M, F172M, N219M and N279N, with one field observed in F154W.
 With the failure of the NUV detector partway through the survey, only about half of the fields were observed in NUV (see Table 1 of \citealt{2020ApJS..247...47L}). 
The field 8 observation was missed in the initial survey. 
The nominal exposure for each observation was 10,000 s, but in practise the exposure times varied from $\simeq$2000 s to 10,000 s. 

The astrometry was calibrated using star positions from Gaia using the CCDLAB UVIT processing software \citep{2017PASP..129k5002P}. 
Because of the differing exposures the sensitivity in each filter, the limiting AB magnitudes varied slightly between fields. 
The limiting AB magnitudes are given by the histograms in the top panel of Fig. 7 in \citet{2020ApJS..247...47L}: F148W- 23.2; F154W- 23.4; F169M- 23.4; F172M- 22.2; N219M- 22.4; N279N- 21.4.
The detector is a photon counting detector: a photocathode followed by a microchannel plate amplifier, then a phosphor to convert the electron cloud into a light shower which is detected by a CMOS array. 
Cosmic ray tracks are removed using the CCDLAB software, resulting in a very low background where noise is dominated by Poisson statistics.  
E.g. the F148W image with observation in 2016 Oct. had $1.04\times10^7$ photon counts, over a circular field covering $1.32\times10^7$ pixels. 
This gives a mean counts and signal-to-noise over a 1 arcminute square region of  $1.63\times10^4$ and 128, respectively.  
 
 \begin{deluxetable*}{cccccccc}
\tablecaption{UVIT Observations for Field 1\label{tab:obs}}
\tablewidth{700pt} 
\tabletypesize{\scriptsize}
    \tablehead{\colhead{UVIT Filter} &  \colhead{Observation (BJD$^a$)} &  \colhead{Exposure time(s)} & \colhead{Counts} &\colhead{mean S/N per \arcmin$^2$} \\ }
\startdata
F148W& A: 2457671 & 7736 & $1.04\times10^7$ & \\
F148W& B: 2458805 & 17191 & $2.39\times10^7$ & \\
F148W& C: 2459183 & 12632 & $1.75\times10^7$ &\\
F148W& A+B+C$^b$ & 38299 & $5.27\times10^7$ & 288 \\
F169M& 2458804 & 10427 & $9.50\times10^6$ & 122 \\
F172M& A: 2457672 & 3612 & $1.63\times10^6$ & \\
F172M& B: 2458805 & 18057 & $8.74\times10^6$ &\\
F172M& A+B$^b$ & 21789 & $1.05\times10^7$ & 128 \\
N219M& 2457671 & 7781 & $6.05\times10^6$ & 97.4 \\
N279N& 2457672 & 3627 & $5.70\times10^6$ & 94.6 \\
\enddata
\tablecomments{a: Barycentric Julian Date. b: Combined images for F148W and F172M were used in the analysis here.}
\end{deluxetable*}

Additional UVIT observations of the bulge (Field 1) in F148W, F169M and F172M filters took place $\sim$3.1 years later.
Table 1 of \citet{2021AJ....161..215L} lists both sets of observations of the bulge field (Field 1).
A third observation of Field 1 was obtained a year later.
Table~\ref{tab:obs} gives the filters, observation dates, exposure times, and mean signal-to-noise (S/N) for a 1 arcminute square region of each image used in this analysis.
All images were produced using CCDLAB \citep{2017PASP..129k5002P}, including combining
the three F148W observations and two F172M observations to produce single deeper exposures in those filters.

\subsection{M31 bulge modeling using GALFIT} \label{sec:galfit}

Two-dimensional modelling of the M31 bulge and nucleus light distribution in NUV and FUV was carried out
using the galaxy image-fitting program GALFIT developed by \citet{2002AJ....124..266P}.   
GALFIT is capable of modelling well resolved galaxy images with multiple components, which enables extraction of galactic substructures that are otherwise hidden. 
GALFIT was designed to model bulges, disks (including spiral arm structure) and halos of galaxies by fitting their images. 
Various functional forms for the different components, and linear combinations of those functions, can be used. 
The functions are elliptical (functions of an elliptical  radial coordinate), which has 4 parameters: 2 for centre ($x_c, y_c$), and 2 for the ellipse  (ellipticity $q$ and major axis position angle $\theta_{PA}$).
Generally a uniform background level of the image is included (1 parameter). 

GALFIT carries out a model fit by the creation of model images, convolution with the PSF and subsequent comparison of the model with the data. 
The noise image can be calculated by GALFIT for the case that the image data is photon counts (Poisson statistics), which is the case for the UVIT images.  
The best-fit model parameters are found by minimizing $\chi^2$ using the Levenberg-Marquardt downhill gradient  method. 
The output of GALFIT includes the best-fit parameters, the model image and the residuals between the 
model and the data.

\subsubsection{Summary of GALFIT Radial Profiles}\label{sec:profiles}

The Sersic function is defined as: 
\begin{equation}
\Sigma(r) = \Sigma_e exp \left[-\kappa\left(\left(\frac{r}{r_e}\right)^{1/n}-1\right)\right]
\end{equation}
The Sersic function has 7 parameters, including the elliptical parameters: $x_c, y_c, m_{tot}, r_e, n, q, \theta_{PA}$, 
with 
$m_{tot}$ the integrated magnitude, $r_e$ the effective radius, and $n$ the Sersic index. 
The Exponential profile has the form:
\begin{equation}
\Sigma(r) = \Sigma_0 exp \left(\frac{-r}{r_s}\right)  
\end{equation}
with 6 parameters: $x_c, y_c, m_{tot}, r_s, q, \theta_{PA}$. 
The Gaussian profile has the form:
\begin{equation}
\Sigma(r) = \Sigma_0 exp \left(\frac{-r^2}{2\sigma^2}\right) 
\end{equation}
with 6 parameters: $x_c, y_c, m_{tot}, FWHM, q, \theta_{PA}$, with $FWHM = 2.354\sigma$. 
The Moffat profile is given by: 
\begin{equation}
\Sigma(r) = \frac{\Sigma_0}{[1+(r/r_d)^2]^n}
\end{equation}
with $r_d=\frac{FWHM}{2\sqrt{2^{1/n}-1}}$.
The 7 parameters are $x_c$, $y_c$, $m_{tot}$, $FWHM$, concentration index $n$, $q$ and $\theta_{PA}$. 
The Nuker profile takes the form:
\begin{equation}
I(r) = I_b  2^{\frac{\beta-\gamma}{\alpha}}\left(\frac{r}{r_b}\right)^{-\gamma}\left[1+\left(\frac{r}{r_b}\right)^{\alpha}\right]^{\frac{\gamma-\beta}{\alpha}}
\end{equation}
The Nuker profile has a total of nine free parameters:
$x_c$ ,$y_c$, $\mu_b$, $r_b$, $\alpha$, $\beta$, $\gamma$, $q$, $\theta_{PA}$, with  $\mu_b$ the surface brightness at radius $r_b$.
Additional available functions are described in \citet{2002AJ....124..266P} and the GALFIT documentation on the GALFIT web page\footnote{The GALFIT home webpage is at https://users.obs.carnegiescience.edu/peng/work/galfit/galfit.html.}.

\subsubsection{UVIT point-spread function (PSF)} \label{sec:psf}

To account for telescope optics and, for ground-based telescopes, atmospheric seeing, GALFIT convolves the model image with a PSF image.
To extract the PSF image for each UVIT filter, the following process was implemented.
For the UVIT image in each filter, several isolated point source stars were identified.
From these we chose one where the star was best centred in the 5 by 5 array of pixels around the star.
CCDLAB produces images with 0.4168\arcsec pixels, which was designed to oversample the instrument resolution of $\simeq$1\arcsec by a factor of $\sim$2.5. 
Because the PSF has extended wings \citep{2017AJ....154..128T} we selected a sub-image
of 55 by 55 pixels centered on the point source. Because there is diffuse light in M31, e.g. from the 
bulge and disk, we subtracted the mean of the four 7 by 7 pixel corners from the image to produce
the diffuse background subtracted PSF.
The GALFIT modelling for each UVIT filter image of M31 utilized the corresponding filter’s PSF fits image. 

\subsubsection{Modelling the M31 bulge structure in the five UVIT filters} \label{sec:modelling}

\begin{figure*}
\gridline{\fig{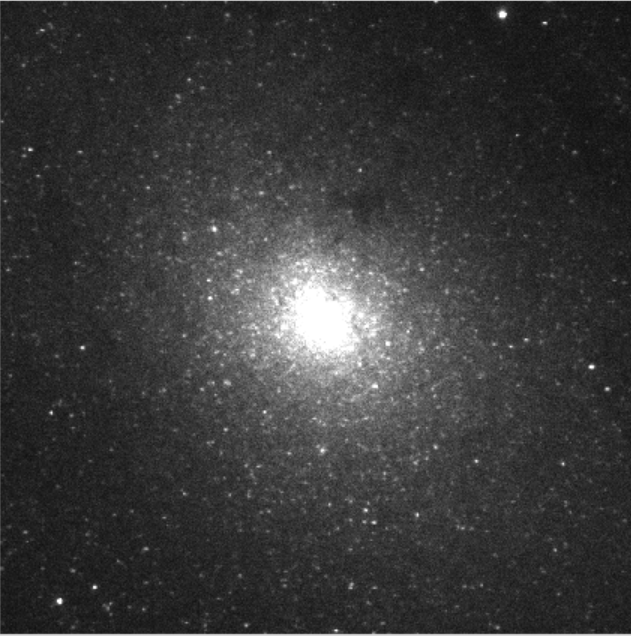}{0.39\textwidth}{(a)}
          \fig{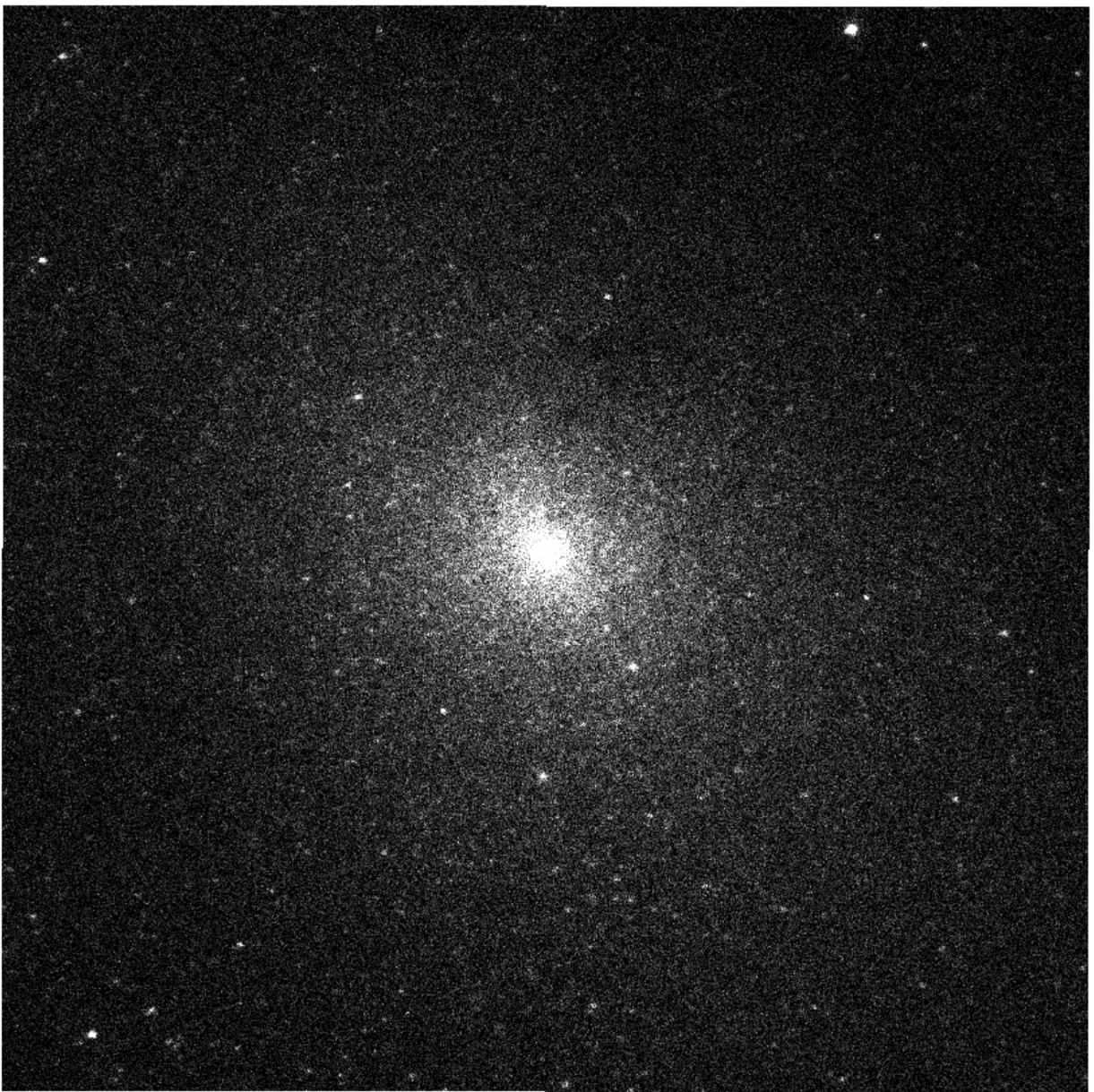}{0.39\textwidth}{(b)}
          }
\gridline{ \fig{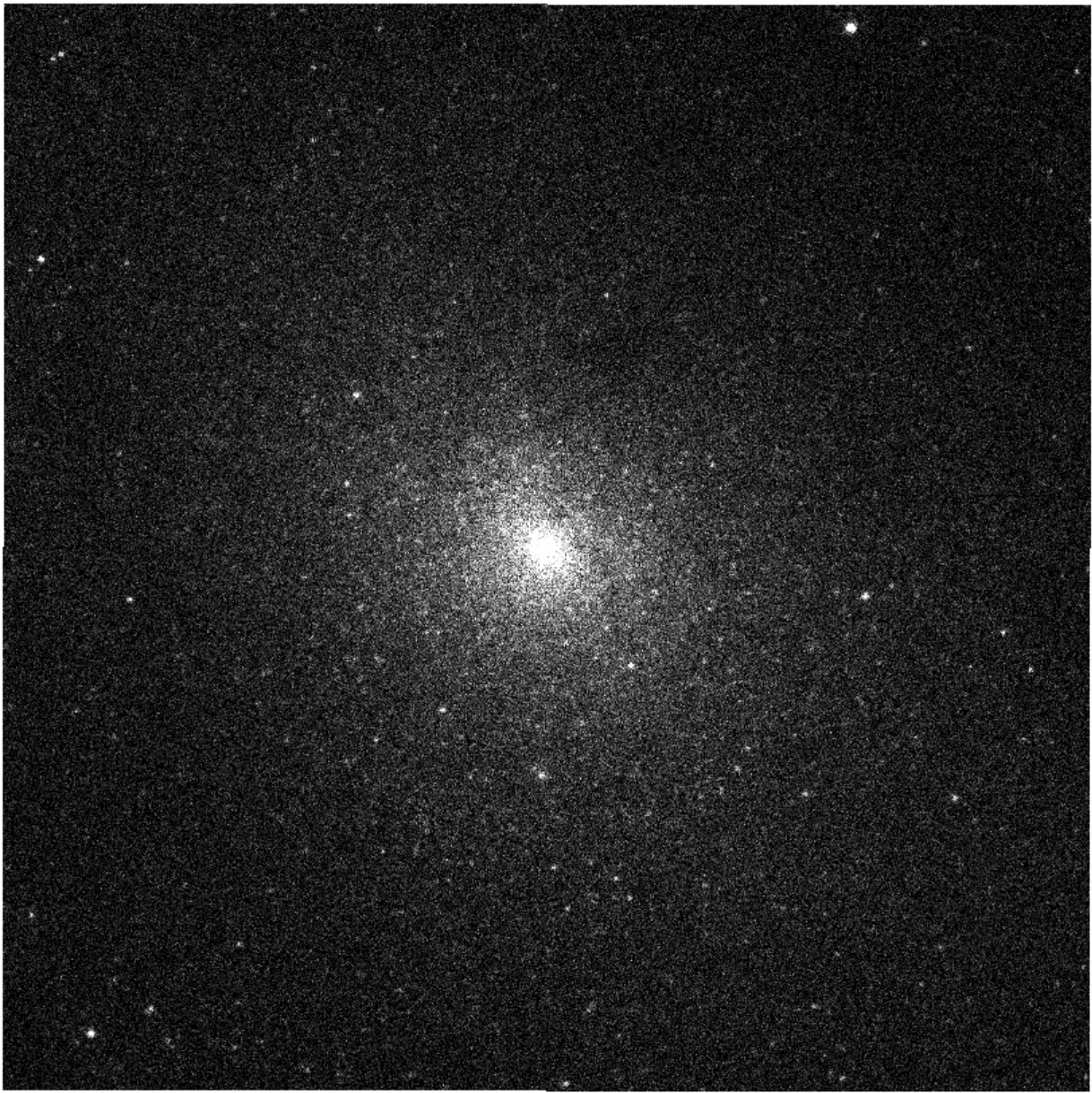}{0.39\textwidth}{(c)}
             \fig{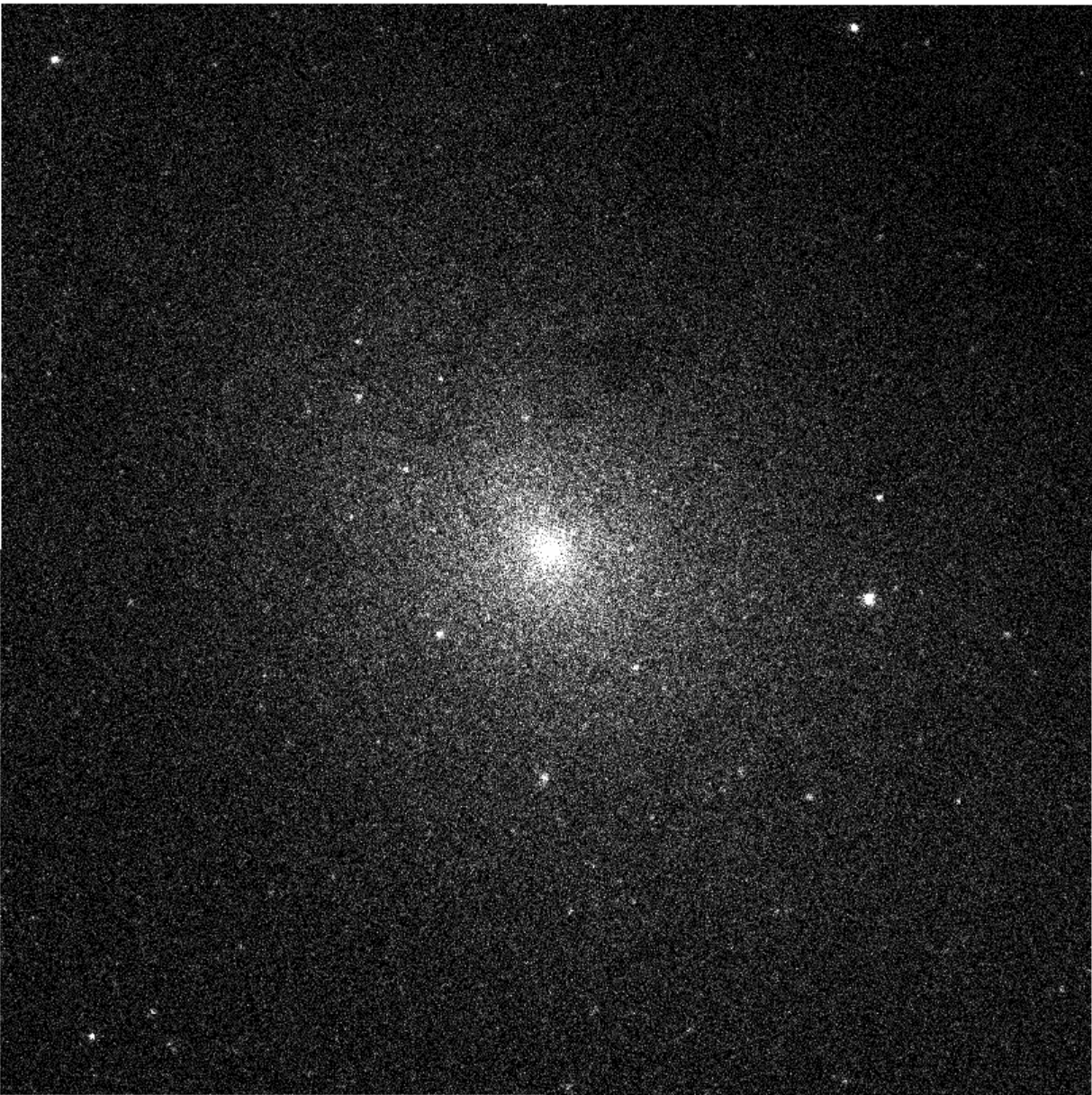}{0.39\textwidth}{(d)}
          }
          \gridline{
          \fig{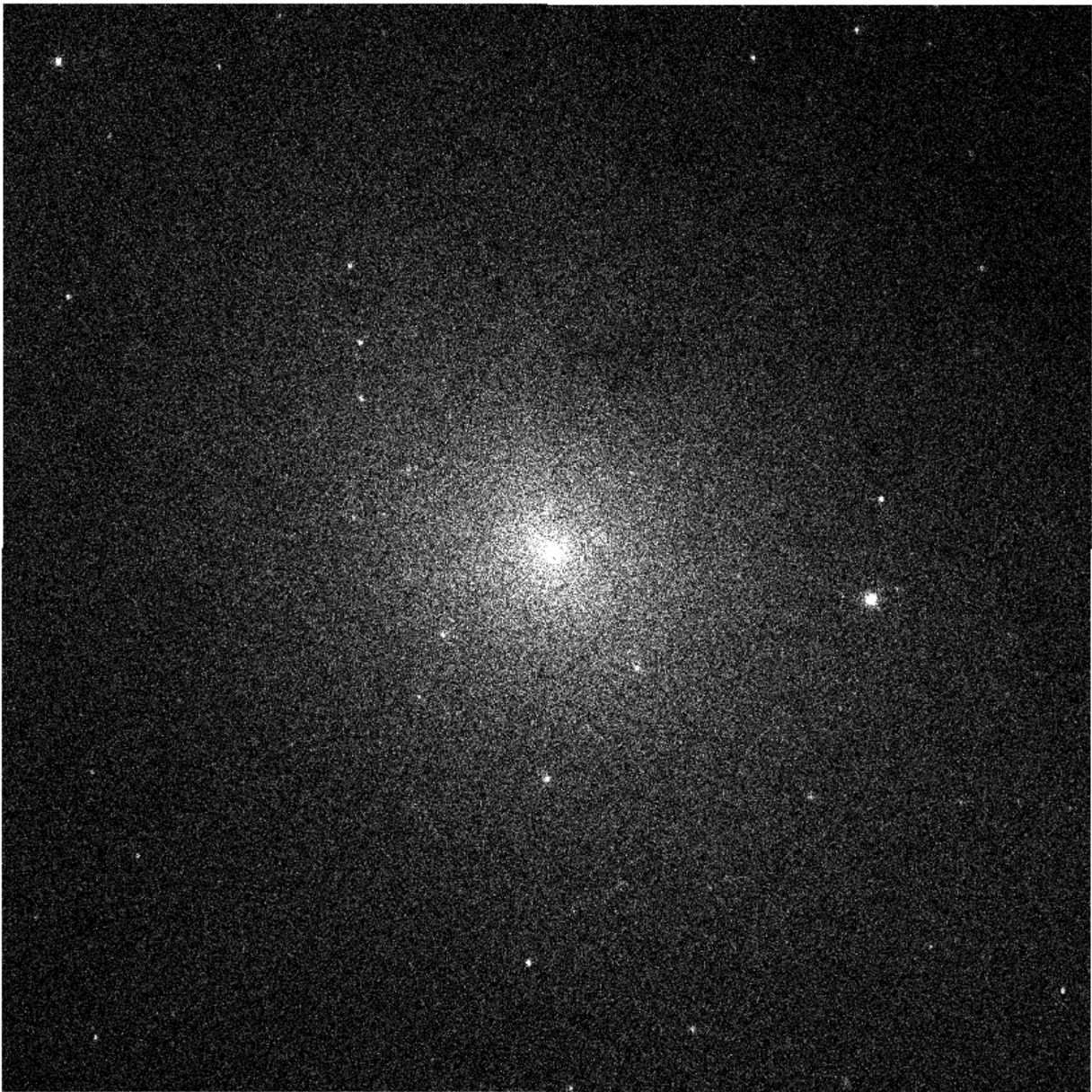}{0.39\textwidth}{(e)}
          \fig{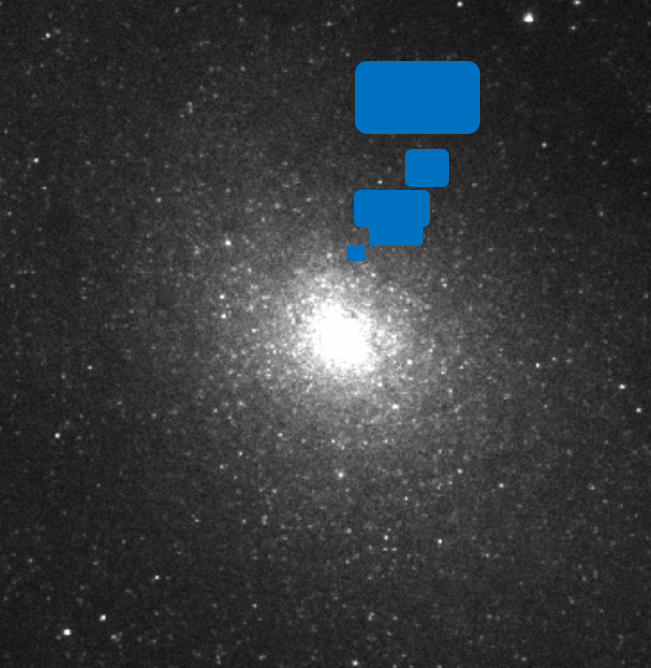}{0.39\textwidth}{(f)}
          }
\caption{UVIT 1001 by 1001 pixel (417\arcsec by 417\arcsec) images of the M31 bulge in: (a) F148W filter; (b) F169M filter;
 (c) F172M filter; (d) N219M filter; (e) N279N filter; (f) F148W filter with dust mask. 
  The dust mask used to omit the dust cloud features is marked in blue in the F148W image.  The pixel size is 0.4168\arcsec. 1\arcsec at M31's distance is 3.81 pc.
\label{fig:UVITposn}}
    \figurenum{1}
\end{figure*}

To model as much of the bulge light as possible while avoiding the light from the disk and inner spiral arms, the central 1001 by 1001 pixel (417.2\arcsec by 417.2\arcsec) area of Field 1 was chosen. 
We visually inspected larger regions (1101 and 1201 pixels) but could see some light from the spiral arms in those images at the corners so rejected use of larger regions.  
The image of the chosen region is shown in Figure~\ref{fig:UVITposn} for the five UVIT FUV and NUV filters: F148W, F169M, F172M, N219M and N279N. 

An initial array of GALFIT tests was performed with the goal of determining the approximate fit qualities of 
the different radial profile functions of GALFIT 
(we tested the five functions listed in Section~\ref{sec:profiles}) 
for both the bulge and nucleus using the F148W image.
The model included the three brightest stars in the chosen 1001 by 1001 pixel area\footnote{The fourth brightest star had significantly less flux.}. 
The Sersic, Gaussian, Moffat, Nuker and PSF functions in GALFIT were used as model components in trials
where up to two components were used to fit the bulge region.
All models components were elliptical \citep{2002AJ....124..266P}, because galactic components (e.g., bulge, disk, and halo) projected onto a 2-dimensional image are elliptical in shape. 

Next we carried out model fits to test the constancy of the model parameters vs. wavelength. 
Three separate series of tests with a Sersic bulge and a Moffat nucleus were performed with: 
i) all free parameters; ii) a constant effective radius; and iii) a constant Sersic index. 
After performing an analysis of the parameters vs. wavelength for each of the three tests, 
we conclude that none of the parameters are consistent with being constant vs. wavelength.
Thus further modeling requires Sersic index, FWHM or concentration index as free parameters for each UVIT filter.

\subsubsection{Dust mask} \label{sec:dustmask}

Dust cloud absorption features are visible northwest of the centre of the bulge in the UVIT images
(see the panels for the various filters in Figure~\ref{fig:UVITposn}, panels a to e).
We excluded the areas with strong dust cloud features from the fits by using a mask
because our purpose was to fit the bulge structure and not the dust clouds. 
We used a python script to generate a list of pixels that correspond to the coordinates of the pixels affected by the main dust cloud regions visible in the bulge. 
The F148W image was used to identify the dust clouds owing to its better signal to noise ratio compared to the images taken with the other four filters. 
The mask is shown in Figure~\ref{fig:UVITposn} (panel f), and was used for the model fits for all five UVIT filters.

\subsubsection{Bright Stars and Background} \label{sec:background}

\begin{deluxetable*}{cccccccc}
\tablecaption{Comparison of Fits for Models With and Without Bright Stars for F148W Image\label{tab:stars}}
\tablewidth{700pt} 
\tabletypesize{\scriptsize}
    \tablehead{\colhead{Bulge/Nucleus} &  \colhead{Star1} & \colhead{Star2} & \colhead{Star3} & \colhead{$\chi^2$} & \colhead{No. free parameters} \\ }
\startdata
Sersic& - & - & - & 728365 & 8 \\
Sersic, PSF&-  &-  & - & 723354  &  11 \\
Sersic, Gauss$^a$ &-  & - & - &719700   &  14 \\
Sersic, Gauss& Gauss & - & - & 704212 & 20 \\
Sersic, Gauss& Gauss & Gauss & - & 702767  & 26 \\
Sersic, Gauss& Gauss & Gauss & Gauss & 698236  & 32 \\
\enddata
\tablecomments{a. Gauss stands for a Gaussian function.}
\end{deluxetable*}

In addition to the stellar light from the bulge there are several bright foreground stars in the image.
We test adding a different numbers of stars in the model for the bulge.
Rows 4 through 6 of Table~\ref{tab:stars} show the $\chi^2$ for model fits including the bulge (Sersic plus Gaussian) plus 1, 2 or 3 stars 
for comparison to the $\chi^2$ for a bulge model with a Sersic plus Gaussian and no stars (row 3).  
Including brightest star, 2 brightest stars or 3 brightest stars successively improves the 
$\chi^2$ by $\sim15000$, $\sim1500$ and $\sim4500$ compared to no stars\footnote{The reason that these values are not monotonically decreasing with decreasing brightness of the stars is because the location of the stars relative to the bulge centre has a significant effect.}. 
 
We did not include more stars because the improvement in fit quality ($\chi^2$) was significantly less than for the three brightest stars. 
For all of the fits presented here the background and the positions and magnitudes of the three brightest stars are free parameters (3 per star), to give the number of background plus star parameters of 10.

\subsubsection{Nucleus mask}\label{sec:nucmask}

The nucleus of M31 has a complex structure as shown by \citet{2002AJ....124..294P}, which includes
the compact nucleus and the nuclear bulge surrounding it.
The compact nucleus was resolved into double components by \cite{1993AJ....106.1436L} with HST (Hubble Space Telescope) Wide Field and Planetary Camera (WFPC) observations.
The HST WPPC2 F555W (V band) image was analyzed by \citet{2002AJ....124..294P} using GALFIT two-dimensional fitting of the central 5\arcsec by 5\arcsec area.  
\cite{1999ApJ...522..772K} show that the double nuclei in V band (called P1 and P2) are the two ends of an eccentric disk of stars orbiting the central black hole which has mass $\sim3\times10^7$M$_{\odot}$.
\citet{2002AJ....124..294P} finds that the central region is well modelled four physical components: 
P1, P2, the M31 elliptical ($q\simeq$ 0.81) bulge and a spherical nuclear bulge ($R_e\simeq$3.2\arcsec). 

In order to fit the brightness distribution of the bulge only, we exclude from the fit the emission associated with the nucleus (compact nucleus plus nuclear bulge). 
For this purpose a series of GALFIT pixel masks of square shape and varying size, centered on the nucleus were created. 
The best fit $\chi^2$ was given by the nucleus pixel mask of dimensions 41 by 41 pixels (17\arcsec \ by 17\arcsec). 
 Table~\ref{tab:nucmask} shows the results of tests conducted to verify whether masking the nuclear region improves the fit of the bulge. 
The $\chi^2$ is reduced at high significance (several 10's of $\sigma$, \citealt{1992nrfa.book.....P})\footnote{The reduction in $\chi^2$ is much larger than the reduction in degrees of freedom (1681 for the 41 by 41 mask).}, indicating that there is a separate component (the nucleus) smaller than the 41 by 41 mask.
 The fact that the bulge region is fit significantly better with the nuclear mask for the tested cases (1 Sersic, 2 Sersic and 3 Sersic functions) is a strong indicator that the nuclear region has additional components, in agreement with the results of  \citet{2002AJ....124..294P}.
Hereafter, when fitting the `bulge only' the nucleus mask is included to omit the nuclear region 
(see Section~\ref{sec:bulgemodel}).  
When fitting the bulge plus nucleus the nucleus mask is omitted, to include the nuclear region  
 (see Section~\ref{sec:nucmodel}).

\begin{deluxetable*}{ccccc}
\tablecaption{Comparison$^a$ of Fits for Models With and Without Nucleus Mask for F148W Image\label{tab:nucmask}}
\tablewidth{700pt} 
\tabletypesize{\scriptsize}
    \tablehead{\colhead{Bulge Components} & \colhead{$\chi^2$ with Mask} & \colhead{$\chi^2$ no Mask} & \colhead{$\Delta\chi^2$} & \colhead{No. free parameters} \\ }
\startdata
Sersic & 685595 & 698695 & -13099  & 8 \\
Two Sersic & 673491 & 683950 & -10459  & 15 \\
Three Sersic & 670345 & 682471 & -12126  & 22 \\
\enddata
\tablecomments{a: model fits include the dust mask. 
}
\end{deluxetable*}

\section{Results}\label{sec:results}

\subsection{Model for the bulge without nuclear region}\label{sec:bulgemodel}

As determined by the tests above, we include the dust mask, the nucleus mask and the three brightest stars 
to find a model for the bulge only, omitting the complex nuclear region \citep{2002AJ....124..294P}. 
The bulge model was gradually increased in complexity by starting with a single Sersic function, 
then adding additional Sersic functions to improve the model fit\footnote{Other functions besides Sersic were tested, including Moffat and Gaussian, but Sersic functions gave the best results so are presented here.}.

\begin{deluxetable*}{ccccc}
\tablecaption{Comparison$^a$ of Fits for Models With and Without Fixed Primary Sersic  Index ($n=2$) for F148W Image\label{tab:nfree}}
\tablewidth{700pt} 
\tabletypesize{\scriptsize}
    \tablehead{\colhead{Bulge Components} & \colhead{$\chi^2$ ($n=2$)} & \colhead{$\chi^2$ ($n$ free)} & \colhead{$\Delta\chi^2$} & \colhead{No. free parameters$^{b}$}\\ }
\startdata
Sersic & 686199 &  685595  & -604 & 7, 8 \\
Two Sersic  & 673491 &  673491& -447& 14, 15 \\
Three Sersic &  670736  &  670345 & -391& 21, 22 \\
\enddata
\tablecomments{a: model fits include the dust mask and nucleus mask. b: the fits with n fixed have one fewer parameters.
}
\end{deluxetable*}
 
A Sersic index $n\sim2$ has been previously suggested for M31 (e.g. \citealt{2011ApJ...739...20C}). 
Thus we compare the models in Table~\ref{tab:nfree} with the primary Sersic function 
(defined as the one with the brightest magnitude) fixed at $n=2$ to models with free index. 
Comparing similar complexity fits (row 1 with row 4, row 2 with row 5, row 3 with row 6) we
see that the improvements in $\chi^2$ by adding one free parameter (the free primary Sersic index)
are 604, 447 and 391, respectively. 
These values are all highly significant decreases in $\chi^2$, showing
that the free primary Sersic index models are better (by 20 to 25 $\sigma$ for change in 1 parameter,  \citealt{1992nrfa.book.....P}) than the $n=2$ cases.
Table~\ref{tab:3sersic} lists the parameters for the 3 Sersic models for the bulge (rows 3 and 6 of Table 2). 
The model with primary Sersic index fixed at $n=2$ has an unrealistically large effective radius: 
2240\arcsec at the distance of M31 is 8.5 kpc, which can be compared to values of 1.0 kpc from \cite{2011ApJ...739...20C} and 0.8 kpc from \cite{2013ApJ...779..103D}. This indicates  
that the $n=2$ fit is unrealistic.
Hereafter we allow the Sersic indices of the Sersic components to be free parameters.

\begin{deluxetable*}{ccccccc}
\tablecaption{Comparison of 3-Sersic Bulge Model Parameters with Primary Sersic (Sersic1) Index Free and Fixed ($n=2$) \label{tab:3sersic}}
\tablewidth{700pt} 
\tabletypesize{\scriptsize}
 \tablehead{\colhead{Model} & \colhead{Component} & \colhead{F148W AB Magnitude} & \colhead{Effective Radius} & & \colhead{Sersic Index} \\ }
\startdata
& &   &  (\arcsec) & (kpc) &  & \\ 
\hline
Primary Sersic Index Free & Sersic1&13.30 &86.2 & 0.328 &1.48 \\
&Sersic2&14.67 &197 & 0.750 &0.31 \\
& Sersic3&12.78 &393 & 1.50 &1.12 \\ 
Primary Sersic Index $n=2$  &Sersic1&9.83 &2240 & 8.53 &2.00 \\
&Sersic2&13.96 &70.0 & 0.266 &1.39 \\
&Sersic3&11.22 &1100 & 4.19 &2.65 \\
\enddata
\end{deluxetable*}

Having determined that Sersic index should be a free parameter, we present the series of fits to the bulge only (using the nucleus mask) of increasing complexity. 
Table~\ref{tab:basic} shows the results (rows 1 to 4) for the bulge modelled using one to four Sersic functions, with each 
Sersic component having 7 free parameters.
The decreases in $\chi^2$ by adding the second, third and fourth Sersic functions are $\simeq$12000, 
$\simeq$3100 and  $\simeq$1800, respectively. 
These decreases are highly significant for addition of 7 extra parameters for each comparison:
E.g. extrapolating from the table on page 815 of \citet{1992nrfa.book.....P}, for 7 parameters of interest, 
a 3$\sigma$ confidence level occurs for an decrease of $\chi^2$ of 22.1, much less than the obtained decreases.  

The maximum number of components we use for modelling the bulge without the nuclear region is four.
The four component fit (row 4) of Table~\ref{tab:basic} gives the best fit of the series that we used.
It is better than the three Sersic fit (row 3) with a $\chi^2$ smaller by 1760 for 7 extra parameters,
which is highly significant \citep{1992nrfa.book.....P}.
 We could have added additional components beyond four before including the central nuclear region. 
Because the next step is fitting the bulge including the nuclear region, we redo the four component model then proceed with adding more functions in the next section below, rather than adding more functions for the fits without the nuclear region.
Inclusion of the central part of image (the nuclear region) in the fits helps to obtain more realistic parameters of the components for the bulge and is essential to determine the components for the nuclear region.

\begin{deluxetable*}{cccccccc}
\tablecaption{Comparison of Fits for Bulge Only Models$^a$ and Bulge Plus Nuclear Region Models$^b$ for F148W Image\label{tab:basic}}
\tablewidth{700pt} 
\tabletypesize{\scriptsize}
    \tablehead{\colhead{Fit No.} &\colhead{Bulge} &  \colhead{Nuclear Bulge$^c$} & \colhead{Nucleus} & \colhead{$\chi^2$} & \colhead{No. of Parameters} & \colhead{Nucleus Mask} \\ } 
\startdata
1&Sersic& - & - & 685595  &   14 &yes\\
2&Two Sersic& - & - & 673491  &  21&yes \\
3&Three Sersic& - & - & 670346& 28 &yes \\ 
4&Four Sersic& - & - & 668586  &  35 &yes \\
\hline 
\hline 
5&Three Sersic& - & PSF & 676565  &  31&no \\
\hline 
6&Sersic& Sersic$(R_e=5.8, n=0.44)$ & - & 690714 &  19&no \\
7&Two Sersic& Sersic$(R_e=5.8, n=0.44)$ & - & 678220 & 26&no \\ 
8&Three Sersic& Sersic$(R_e=5.8, n=0.44)$ & - & 675724  &  33&no \\
\hline 
9&Sersic& Sersic$(R_e=5.8, n=0.44)$ & PSF & 689436  & 22&no \\
10&Two Sersic& Sersic$(R_e=5.8, n=0.44)$ & PSF & 676810  & 29&no \\
11&Three Sersic& Sersic$(R_e=5.8, n=0.44)$ & PSF & 674513  & 36&no \\
12&Four Sersic& Sersic$(R_e=5.8, n=0.44)$ & PSF & 672310  &  43&no \\
\enddata
\tablecomments{a. All fits include the dust mask, background and 3 brightest stars. Bulge only models are the top 4 rows and exclude the nuclear region. b. Rows 5 to 12 include whole image fitting. Row 5 is model for bulge plus compact nucleus, rows 6 to 8 are models for bulge plus nuclear bulge; rows 9 to 12 are models for bulge plus nuclear bulge plus compact nucleus. c. Nuclear bulge Sersic model has effective radius $R_e$ and index $n$ fixed to the values from \citet{2002AJ....124..294P}.}
\end{deluxetable*}

\subsection{Model for the bulge with nuclear region}\label{sec:nucmodel}

M31 has a complex nuclear and bulge morphology so a simple radial luminosity profile is not a sufficient model. 
When fitting the M31 double nucleus using high resolution Hubble Space Telescope data, \citet{2002AJ....124..294P} identified four nucleus components: a UV peak, the dynamic center and two other focal points known as P1 and P2 in the literature. 
That work adds a nuclear bulge component to complete the model for the light distribution. 
In total \citet{2002AJ....124..294P} used six components to achieve an optimal fit. 

\subsubsection{Bulge plus nuclear region model with fixed nuclear parameters}\label{sec:nucfix}

For these fits, the Sersic function for the nuclear bulge has effective radius $R_e$ and index $n$ fixed to the values from \citet{2002AJ....124..294P}.
We use a PSF function for the compact nucleus because it is unresolved with UVIT, although it is complex at the sub-arcsecond scale  \citep{2002AJ....124..294P}.
Table~\ref{tab:basic} shows results (rows 5 to 12) for model fits including the nuclear region (no nucleus mask).
These use various numbers of Sersic functions for the bulge, plus one Sersic for the nuclear bulge and a PSF function for the compact nucleus.

Comparing the different models in Table~\ref{tab:basic}, we find the following. 
First we compare fits 
for the bulge only (top 3 rows in Table~\ref{tab:basic}) with the fits including nucleus (row 5) or nuclear bulge (rows 6 to 8), with the nuclear bulge parameters fixed to the \citep{2002AJ....124..294P} values, except normalization. 
For fits including the pixel mask, the degrees of freedom (dof) is reduced by 1681, from $1002001-N_{par}$ to  $1000320-N_{par}$.
 The three Sersic plus PSF (row 5) for the bulge plus nuclear region (with 1681 more pixels than the bulge with nuclear mask) has higher $\chi^2$ by $\sim$6200 than the three Sersic for the bulge alone (row 3). 
This is significantly higher than the expected increase of 1681, the extra dof introduced by including the 41 by 41 pixel nuclear region. 
This shows that the PSF alone does not provide a good fit to the nuclear region. 
Similarly, comparison of row 1 with row 6, row 2 with row 7,  or 3 with row  8, shows increases of $\chi^2$ of $\simeq$5000 for each case. 
This is much higher than the extra dof introduced, showing the nuclear bulge with fixed parameters does not provide a better fit. 

Next we compare fits to the bulge only (rows 1 to 4) with fits including bulge, nucleus and nuclear bulge (rows 9 to 12).
The one, two, three and four Sersic bulge fits with nucleus and nuclear bulge have increases in $\chi^2$ of 3841, 3319, 4167 and 3724, respectively, over fits without nucleus and nuclear bulge.
This is much greater than the extra dof introduced. 
We conclude that the model with fixed parameter Sersic plus PSF is not a good model for the nuclear region (nucleus plus nuclear bulge).

\begin{figure*}[htbp]
    \centering
    \epsscale{1.}
 \plotone{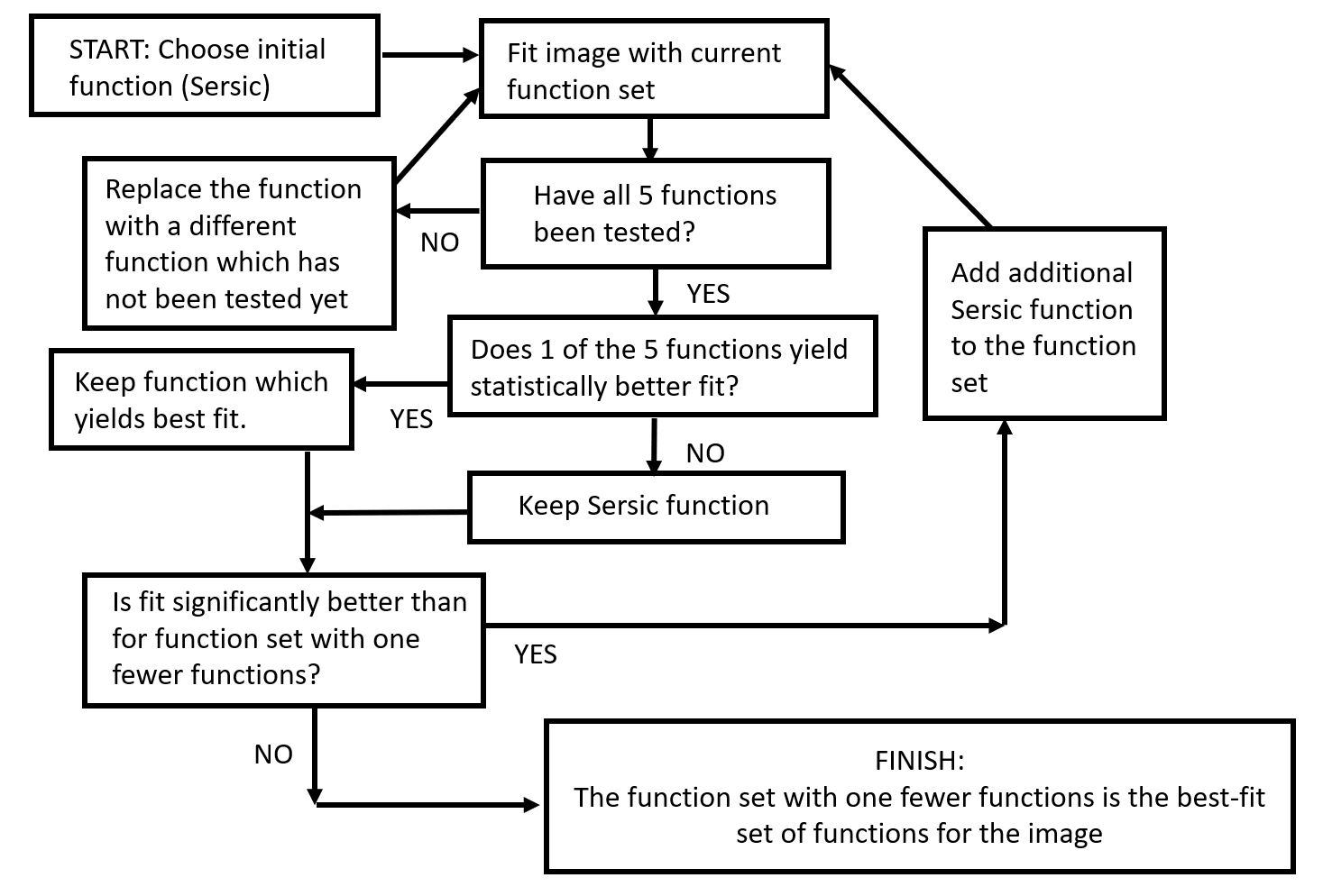}
    \figurenum{2}
    \label{fig:flow}
    \caption{Flowchart for finding best-fit set of functions  to the image of the M31 bulge
    (Sections~\ref{sec:profiles} and \ref{sec:nucfree}). 
}
\end{figure*}

\subsubsection{Bulge plus nuclear region model with free nuclear parameters}\label{sec:nucfree}

For our final set of fits to the bulge plus nuclear region, we use models for the compact nucleus plus nuclear bulge, with nuclear model parameters not tied to the values from \citet{2002AJ....124..294P}.
The nuclear region observed by UVIT in the F148W filter has lower resolution than the HST F555W data, and the wavelength is significantly different, so different stars are prominent in F148W than the HST data. 
The F148W filter extends from 125 to 175 nm \citep{2017AJ....154..128T}, so the bulge and nuclear region structure can be quite different than what was
found at 555 nm by \citet{2002AJ....124..294P}.

The procedure for finding the best-fit model, without fitting beyond what is statistically significant, is outlined in Figure~\ref{fig:flow}.
We start with a single function to model the bulge, including nuclear region, and compare it to fits with the other four functions (see Section~\ref{sec:profiles}). 
The function with the best fit is chosen, then the complexity is increased by adding an additional function to the function set. 
All five functions are tested as the added function, and the function set with the best fit was kept.  
Next we test whether the current function set has a significantly better fit than the best-fit with the function set with one fewer functions. 
If yes, then we test a function set with one additional function, as shown in the flowchart.
If not, the added function is not justified by the data, and the set with one fewer functions is the final best-fit function set. 

Models  with free bulge and nuclear parameters with four to eight components are shown in Table~\ref{tab:BNmodel}, with parameter values and $\chi^2$ values given. 
The model with nine components does not yield a statistically better fit, so is not shown.
Neither do we show the models with three or fewer components, as these are inadequate to fit the bulge plus nuclear region and have significantly worse fits.
The parameter errors are those generated by GALFIT from the fitting process.

We compare the models to comparable models with  
 fixed nuclear parameters, shown in Table~\ref{tab:basic}.
Model 1 in Table 7 has a lower $\chi^2$ by 2479 than Model 8 from Table~\ref{tab:basic} but only 2 more parameters, thus a significantly better fit. 
Comparing Model 1 in Table~\ref{tab:BNmodel} with  Model 11 from Table~\ref{tab:basic}, yields that Model 1 Table~\ref{tab:BNmodel} has a lower $\chi^2$ by 1268 with 1 less parameters.
Thus we conclude that  the free nuclear region parameters result in much better fits, and that the nucleus shape for the 148 nm image is significantly different than for 555 nm. 

We next compare the different models in Table~\ref{tab:BNmodel}.
The models with more components (thus parameters) generally have lower $\chi^2$. 
The $\Delta\chi^2$ values are compared here to determine how much better are the more complex models. 
Model 2, with 41 parameters, has  $\Delta\chi^2=815$ compared to model 1, with 6 extra parameters. 
This is a highly significant improvement in fit \citep{1992nrfa.book.....P}.
Model 3 has  $\Delta\chi^2=1351$ compared to model 2, with 7 extra parameters, again a highly significant improvement in the fit.  
Model 4 has  $\Delta\chi^2=123$ compared to model 3, with 6 extra parameters. 
This is significant improvement in the fit\footnote{$\Delta\chi^2=20.1$ is a 3$\sigma$ improvement for 6 parameters, from \citealt{1992nrfa.book.....P}.}.
Model 5 has $\Delta\chi^2=69$ compared to model 4, with 5 extra parameters, which is a significant improvement in fit. 
We did not proceed with more complex models than Model 5 for two reasons: the improvement in fit was no longer highly significant; and
GALFIT would not converge from almost all the models attempted, likely because the number of parameters was too large.
Thus our final model for the bulge and nuclear region is Model 5 with three components for the bulge and five components for the nuclear region.

\begin{figure*}
\gridline{\fig{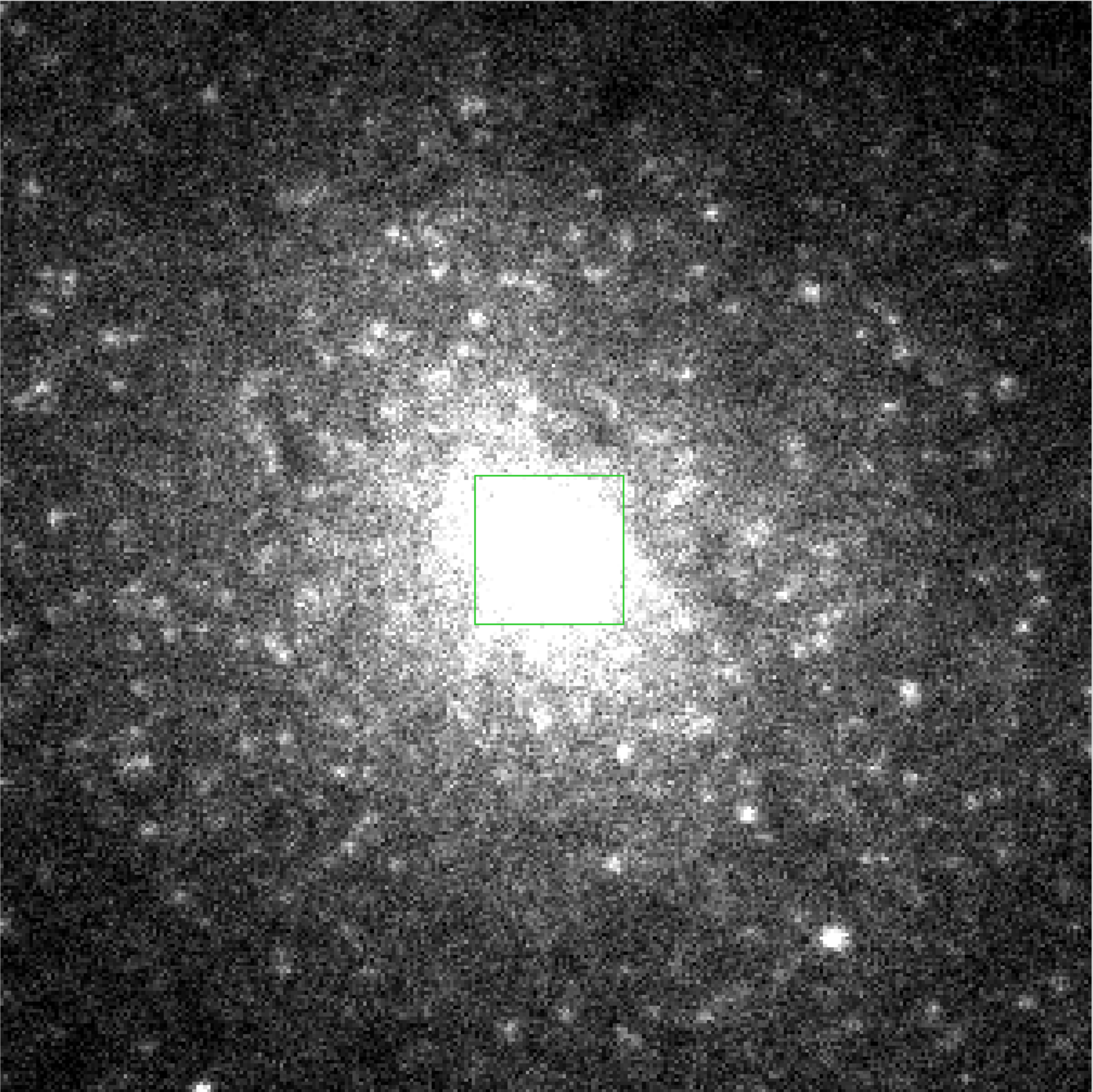}{0.302\textwidth}{(a)}
          \fig{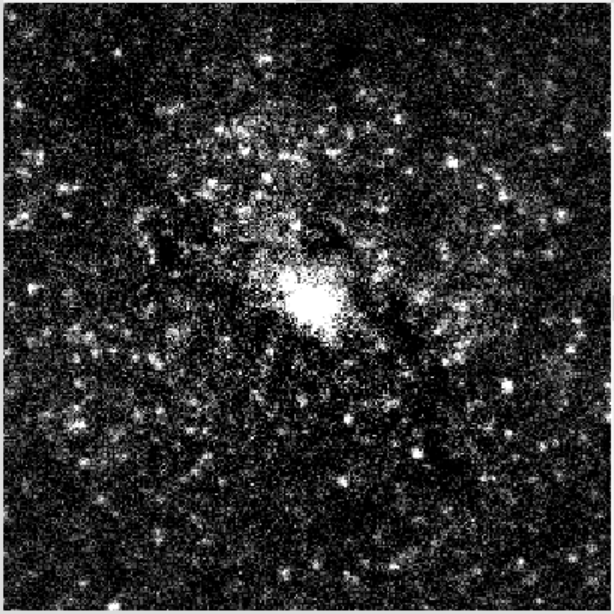}{0.304\textwidth}{(b)}
            \fig{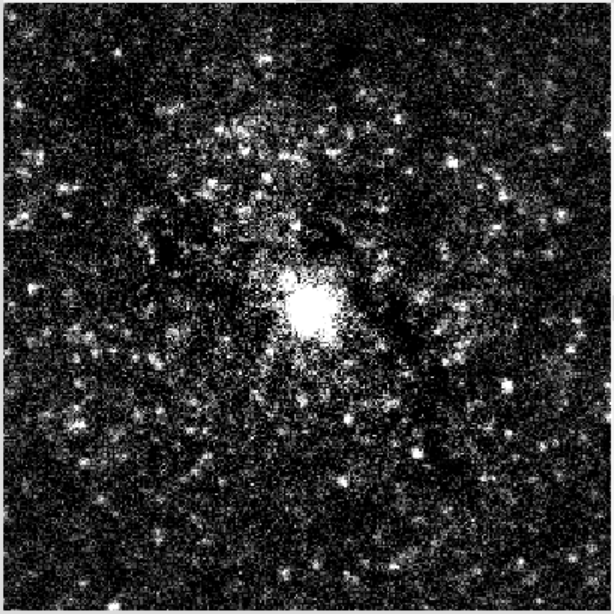}{0.304\textwidth}{(c)}
          }
\gridline{ 
            \fig{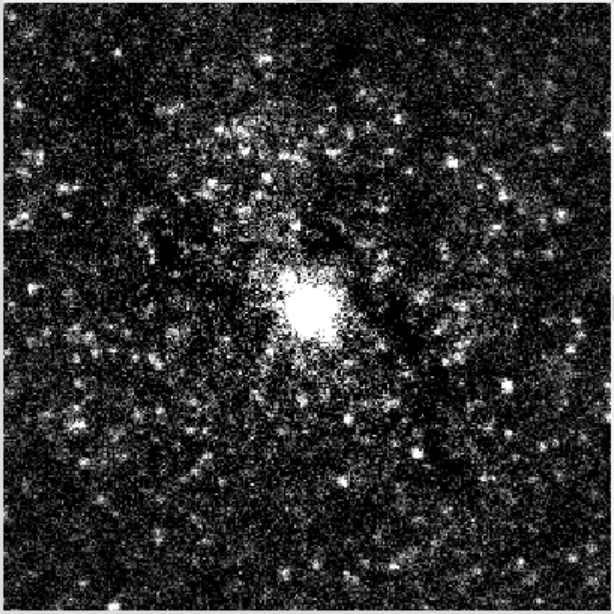}{0.304\textwidth}{(d)}
              \fig{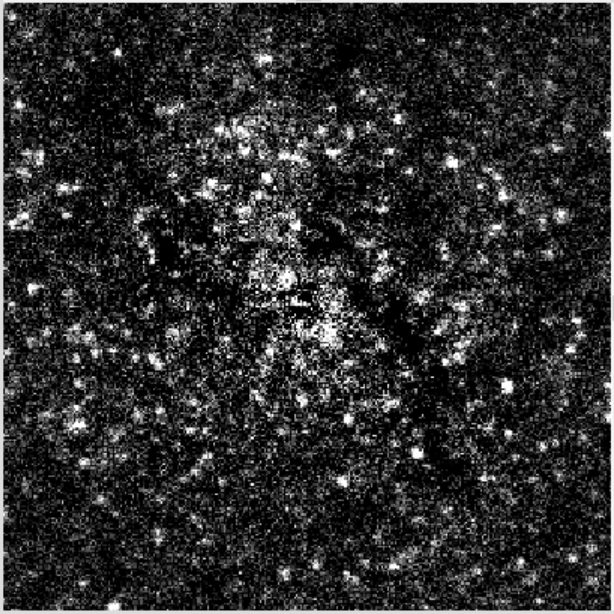}{0.304\textwidth}{(e)}
              \fig{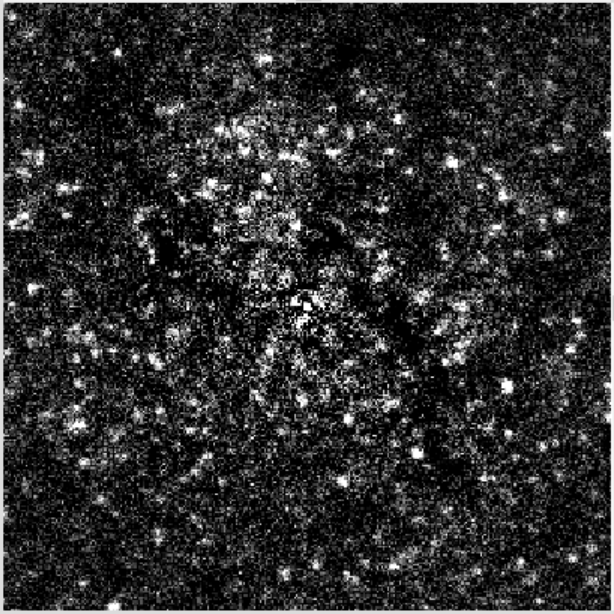}{0.304\textwidth}{(f)}
          }
          \gridline{
             \fig{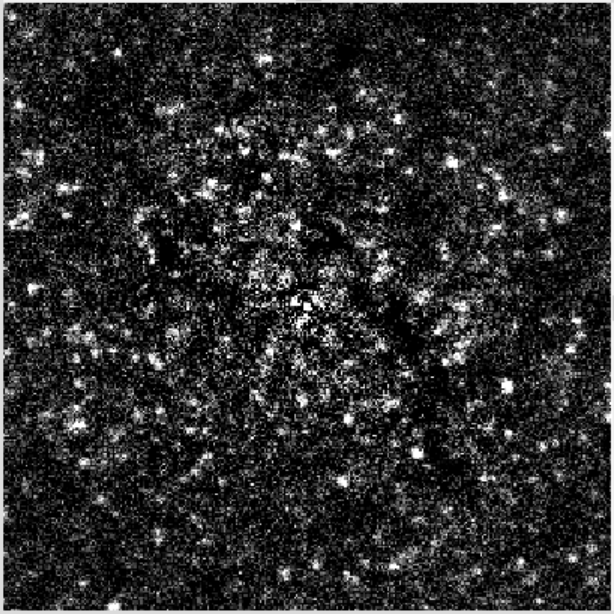}{0.304\textwidth}{(g)}
             \fig{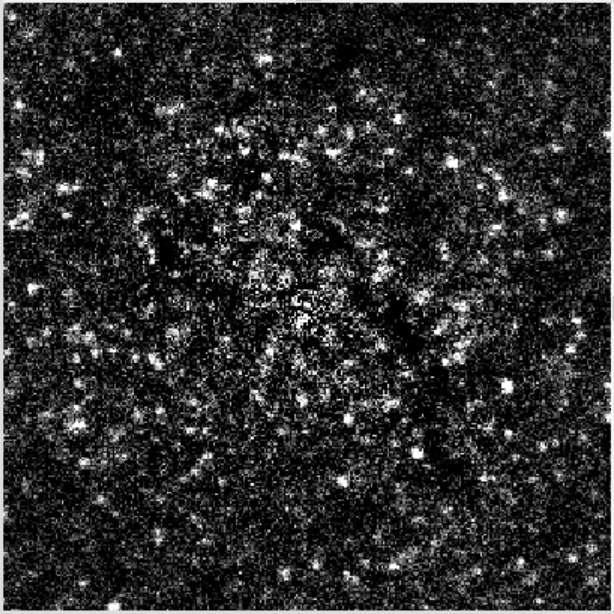}{0.304\textwidth}{(h)}
             \fig{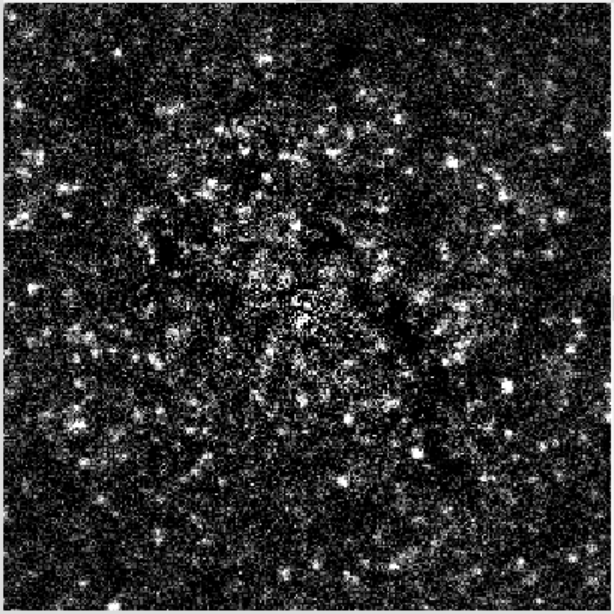}{0.304\textwidth}{(i)}
          }
    \figurenum{3}
    \label{fig:resid}
    \caption{Central part of F148W image (panel (a), top left) and residual images for a series of models 
    (panels (b) to (i), from second image from left in top row to last image in bottom row at lower right). 
    Each image is 125\arcsec by 125\arcsec (475 by 475 pc) centered on the nucleus.
    The greyscale for each image scale is linear from  the image mean -0.5 standard deviations to the 
    image mean +5 standard deviations.  
    Panel (a) shows the nuclear region excluded (green rectangle) for the bulge only fits (b), (c) and (d).
Panels (b), (c) and (d) show the residual images for the best 1, 2 and 3 component fits for the bulge with nuclear region excluded. 
These are the 1, 2 and 3 Sersic models with Sersic index free and which have $\chi^2$ and parameters given in Table~\ref{tab:nfree} and  Table~\ref{tab:3sersic}.
Panels (e) through (i) show the residual images for the best 4, 5, 6, 7 and 8 component fits for the bulge including nuclear region. These models are given in  Table~\ref{tab:BNmodel}.
The residual images are given in order decreasing $\chi^2$ for each set ((b) to (d) fits with nuclear region omitted, and (e) to (i) fits to the entire image. 
 }
\end{figure*}

Residual images are obtained by subtracting model images from the original image.
Figure~\ref{fig:resid} illustrates the process of finding the best combination of functions for the bulge of M31. 
It shows the original F148W observed image (top left) and, in the remaining panels, residual images for successively better models for the central 125\arcsec by 125\arcsec (475 by 475 pc) region centered on the M31 compact nucleus. 

\begin{longrotatetable}
\begin{deluxetable*}{ccccccccccccc}
\tablecaption{Comparison of Parameters for Models$^a$ with Four or More Components for Bulge plus Nuclear Bulge/Nucleus  for F148W Image \label{tab:BNmodel}}
\tablewidth{700pt} 
\tabletypesize{\scriptsize}
    \tablehead{\colhead{Model} &  \colhead{Comp.} & \colhead{Center Offset(\arcsec)} & \colhead{Magnitude} & \colhead{$R_e$(\arcsec or kpc$^b$)} & \colhead{Sersic Index} & \colhead{FWHM(\arcsec or kpc$^b$)} & \colhead{Power Law} & \colhead{Axis Ratio} & \colhead{Position Angle$^b$($^\circ$)} \\ }
\startdata
1. 4 Components & Bulge &  &  &  &  & & & & \\
 Parameter No.: 35 & Sersic & 41.56 $\pm$0.23 & 12.78$\pm$0.12 & 393$\pm$29(1.50$\pm$0.11) & 1.12$\pm$0.04 &  &  & 0.620 $\pm$0.003 & -46.2 $\pm$0.5 \\
$\chi^2$: 673245   & Sersic  &0.76$\pm$0.02&13.30$\pm$0.02&86.2$\pm$1.0(0.328$\pm$0.038)&1.48$\pm$0.01&   &    &0.760$\pm$0.001&-58.1$\pm$0.1\\
   & Sersic & 16.84$\pm$0.71 & 14.67$\pm$0.02 & 196.6$\pm$2.9(0.748$\pm$0.011) & 0.30$\pm$0.01 & & & 0.760$\pm$0.001 & -58.1 $\pm$0.1 \\
& Nucleus &  &  &  &  &  &  &  &  \\
 & Sersic & 0.042$\pm$0.007 & 17.86$\pm$0.01 &2.42(.009) & 1.87$\pm$0.04 &  &  & 1.00$\pm$0.01 & n/a \\
 &  &  &  &  &  &  &  &  &  \\
2. 5 Components & Bulge &  &  &  &  &  &  &  &  \\
 Parameter No.: 41 & Sersic & 41.73$\pm$0.20 & 12.78$\pm$0.12 & 393$\pm$29(1.50$\pm$0.11) & 1.12$\pm$0.03 & &  &  &  \\
$\chi^2$: 672430  & Sersic &0.84$\pm$0.02 & 13.30$\pm$0.02 & 86.2$\pm$1.0(0.328$\pm$0.038) & 1.48$\pm$0.01 &  &  & 0.770 $\pm$0.001 & -58.4 $\pm$0.1
\\
   & Sersic & 16.59 $\pm$0.70 & 14.67$\pm$0.02 & 196.6$\pm$2.9(0.748$\pm$0.011) & 0.30$\pm$0.01 &  &  &  &   \\
& Nucleus &  &  &  &  &    &  &  \\
 & Moffat & 0.28$\pm$0.03 & 18.07$\pm$0.02 &  &  & 11.1$\pm$2.3(0.042$\pm$0.009) & 8.8$\pm$3.0 & 0.70$\pm$0.01  & -34.7 $\pm$1.0 \\
 & Gaussian & 0.023$\pm$0.008 & 19.00$\pm$0.01 &  &  & 2.22$\pm$0.04(0.0084$\pm$0.0002) & & 0.80$\pm$0.02 & -34.7$\pm$1.0 \\
 &  &  &  &  &  &  &  &  \\
3. 6 Components & Bulge &  &  &  &  &  & & &  \\
 Parameter No.: 48 & Sersic &42.88$\pm$0.20 & 12.78$\pm$0.12 & 393$\pm$29(1.50$\pm$0.11) & 1.12$\pm$0.035 & & & 0.63$\pm$0.01 & -46.9$\pm$0.5\\
$\chi^2$: 671077  & Sersic & .98$\pm$0.01 & 13.30$\pm$0.02 & 86.2$\pm$1.0(0.328$\pm$0.038) & 1.48$\pm$0.01 & & & 0.77 $\pm$0.01 & -60.7 $\pm$0.1 \\
   & Sersic & 16.97$\pm$0.71 & 14.67$\pm$0.019 & 196.6$\pm$2.9(0.748$\pm$0.011)& 0.30$\pm$0.01 & & & 0.42$\pm$0.01 & -47.4$\pm$0.2 \\
 & Nucleus &  &  &  &  &  & & & \\
  & Moffat & 0.46$\pm$0.049 & 18.06$\pm$0.019 &  &  & 25.1$\pm$4.9(0.095$\pm$0.018) & 7.87$\pm$2.6 & 0.71$\pm$0.0063 & -38.8$\pm$0.8 \\
  & Moffat & 12.51$\pm$0.15 & 18.74$\pm$0.028 &  &  & 32$\pm$27(0.122$\pm$0.102) & 20.00$\pm$32 & 0.46$\pm$0.011 &  -9.9$\pm$0.9  \\
 & Gaussian & 0.028$\pm$0.008 & 19.03 $\pm$0.01 &  &  & 2.20$\pm$0.04(0.0084$\pm$0.0002) & & 0.77$\pm$0.02 & -73.2$\pm$3.1 \\
 &  &  &  &  &  &  &  &   \\
4. 7 Components & Bulge &  &  &  &  &  &  &  &  \\
 Parameter No.: 54  & Sersic & 42.89$\pm$0.22  & 12.78$\pm$0.12 & 393$\pm$29(1.50$\pm$0.11) & 1.12$\pm$0.035 & &  & 0.63$\pm$0.0025 & -46.85$\pm$0.45 \\
$\chi^2$: 670954  & Sersic &0.9975$\pm$0.015 & 13.30$\pm$0.015 & 86.2$\pm$1.0(0.328$\pm$0.004) & 1.48$\pm$0.011 &  &  & 0.77$\pm$0.01& -60.7$\pm$0.1\\
   & Sersic & 16.97$\pm$0.71 & 14.67$\pm$0.019 & 196.6$\pm$2.9(0.748$\pm$0.011) & 0.30$\pm$0.0058 &  &  & 0.42$\pm$0.0013 &  -47.41$\pm$0.19 \\
 & Nucleus &  &  &  &  &  &  &  &  \\
 & Moffat & 0.154$\pm$0.010 & 18.11$\pm$0.019 &  &  & 11.5$\pm$3.6(0.044$\pm$0.014) & 12.27$\pm$6.9 & 0.67$\pm$0.0072 & -37.4$\pm$1.1  \\ 
& Moffat & 12.69$\pm$0.05 & 18.74$\pm$0.029 &  &  & 30$\pm$25(0.114$\pm$0.095) & 20.00$\pm$32 & 0.51$\pm$0.013 &  -9.70$\pm$1.5  \\ 
& Gaussian & 0.112$\pm$0.024 & 19.41$\pm$0.047 &  &  & 1.88$\pm$0.04(0.0072$\pm$0.0002) & & 0.71$\pm$0.023 &  -48.4$\pm$5.8 \\
& Gaussian & 0.28$\pm$0.15 & 19.95$\pm$0.071 &  &  & 4.34$\pm$0.27(0.0165$\pm$0.0010) & & 0.38$\pm$0.021 &  86.0$\pm$1.9  \\
&  &  &  &  &  &  &  &    \\ 
5. 8 Components & Bulge &  &  &  &  & &  &  &  \\
 Parameter No.: 59  & Sersic & 42.95$\pm$0.20 & 12.78$\pm$0.12 & 393$\pm$29(1.50$\pm$0.11)& 1.12$\pm$0.035 & & & 0.63 $\pm$0.0025 & -46.8$\pm$0.5 \\
$\chi^2$: 670885   & Sersic &0.988$\pm$0.015 & 13.30$\pm$0.015 & 86.2$\pm$1.0(0.328$\pm$0.038) & 1.48$\pm$0.011 & & & 0.77$\pm$0.0008 & -60.8$\pm$0.1\\
   & Sersic & 17.02$\pm$0.71 & 14.67$\pm$0.019 & 196.6$\pm$2.9(0.748$\pm$0.011) & 0.30$\pm$0.0058  & & & 0.42$\pm$0.0013 &  -47.4$\pm$0.2 \\
 & Nucleus &  &  &  &  &  &  &  &  \\
 & Gaussian & 0.075$\pm$0.028 & 18.17$\pm$0.0096 &  &  & 11.73$\pm$0.14(0.0446$\pm$0.0005) & & 0.71$\pm$0.0076 &  -38.9$\pm$1.0  \\
 & Gaussian & 12.58$\pm$0.16 & 18.71$\pm$0.027 &  &  & 30.7$\pm$1.0(0.117$\pm$0.004) & & 0.52$\pm$0.016 & -9.3$\pm$1.2  \\
  & Gaussian & 0.104$\pm$0.009 & 19.30$\pm$0.038 &  &  & 2.01$\pm$0.04(0.0077$\pm$0.0002) & &  0.71$\pm$0.020 & -49.4$\pm$5.3  \\ 
   & Gaussian & 0.32$\pm$0.06 & 20.04$\pm$0.083 &  &  & 4.51$\pm$0.28(0.017$\pm$0.001) & & 0.38$\pm$0.021 & 86.3 $\pm$2.1 \\
  & Sersic & 3.18$\pm$0.16 & 21.95$\pm$0.12 & 5.8(0.022) & 0.24 & & & 0.44$\pm$0.0 &  -2.4$\pm$1.5  \\
\enddata
\tablecomments{a: All fits include the dust mask, background and 3 brightest stars.b. Position angle is counterclockwise from N. b: $R_e$ in kpc is given in round brackets.}
\end{deluxetable*}
\end{longrotatetable}

The best-fit $\chi^2$ values are successively lower for the residual images in successive panels.  
The second panel in the top row is for a single Sersic model for the bulge and no nuclear model (row 1 in  Table~\ref{tab:basic}).
The  single Sersic model, although it accounts for the bulk of the light from the bulge 
(in comparison with the first panel) has a significant residual in the nuclear region (inner 41 by 41 pixel region) 
and thus does not fit the light distribution of bulge plus nuclear region. 
Successive models (residuals in panels 3 and 4, for the 2-Sersic and 3-Sersic models) fit the region outside the 41 by 41 pixel nuclear region and represent the structure of the outer region more accurately, but they do not fit the nuclear region.  
Panels 5 through 9 are for models 1 to 5 in Table~\ref{tab:BNmodel} which include components for the nuclear region. These show increasingly better fits (smaller residuals)
in the nuclear region while maintaining or improving the fit to the outer region. 
The last panel in the bottom row is for our final bulge plus nuclear region model 5 with eight components shown in Table~\ref{tab:BNmodel}.
This model yields no visible artifacts in the nuclear region at the centre nor in the bulge further out.

From the residual images in Figure~\ref{fig:resid}, it is seen that as the fit improves, the residuals in the nuclear region decrease. 
The final model fits well the large scale structure of the bulge and the smaller scale structure of the nuclear bulge and nucleus.  
There remains significant small scale structure in the last image shown. 
The residual structures which are not modeled are seen to consist primarily of dust lanes and a number of small 
(few \arcsec across) bright spots scattered throughout the region. 

\subsection{Bulge/nucleus offset and model asymmetry}\label{sec:offset}

\begin{figure*}
  \figurenum{4}
  \label{fig:asym}
\gridline{\fig{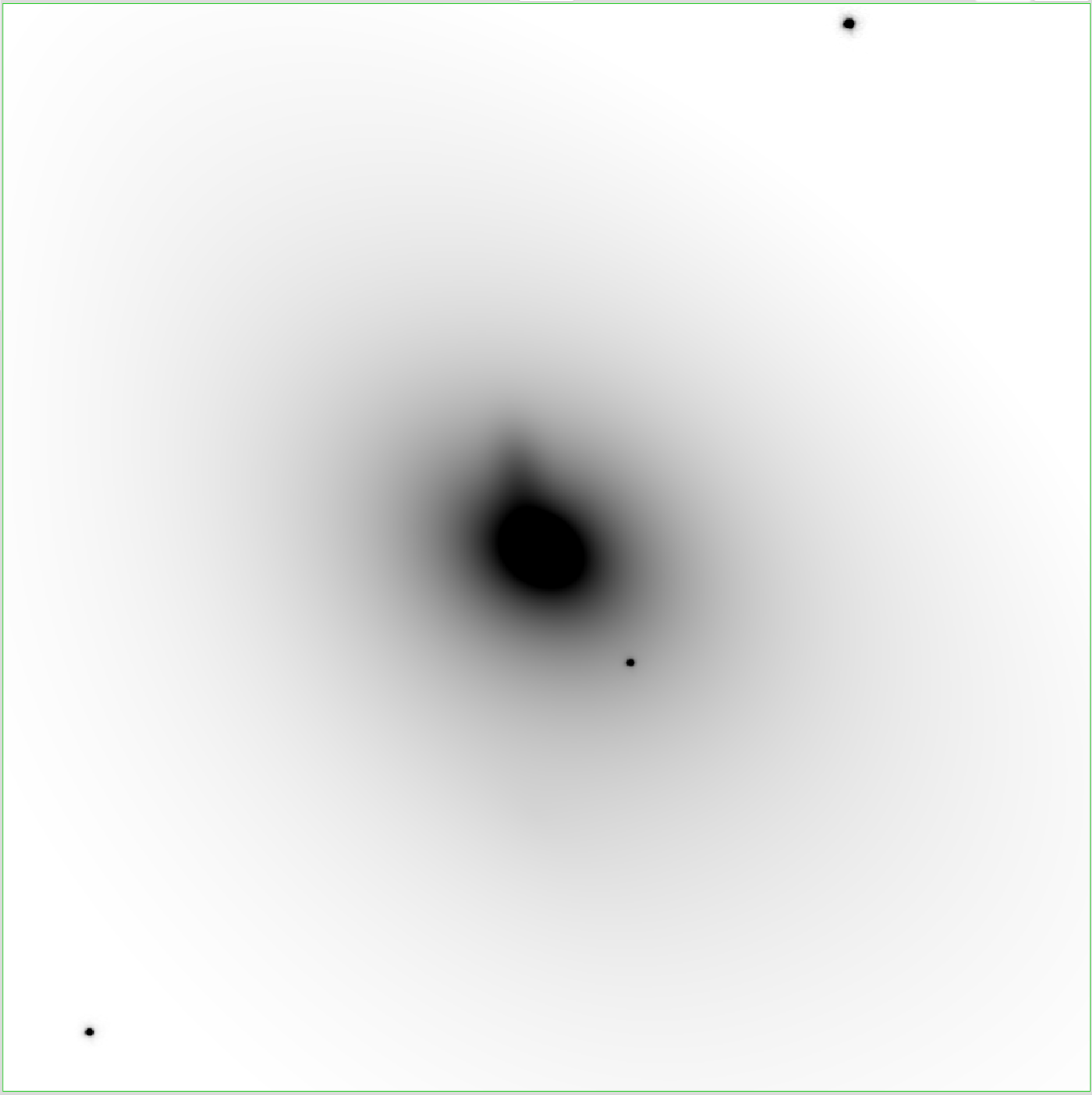}{0.29\textwidth}{(a)}
          \fig{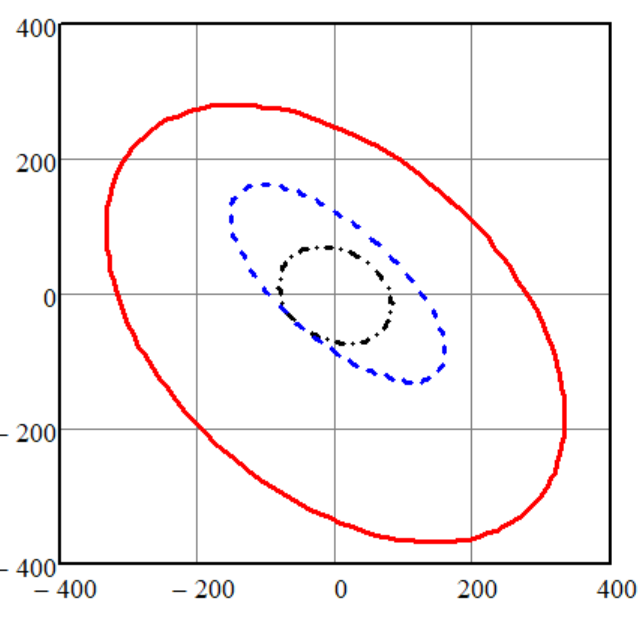}{0.315\textwidth}{(b)} 
          \fig{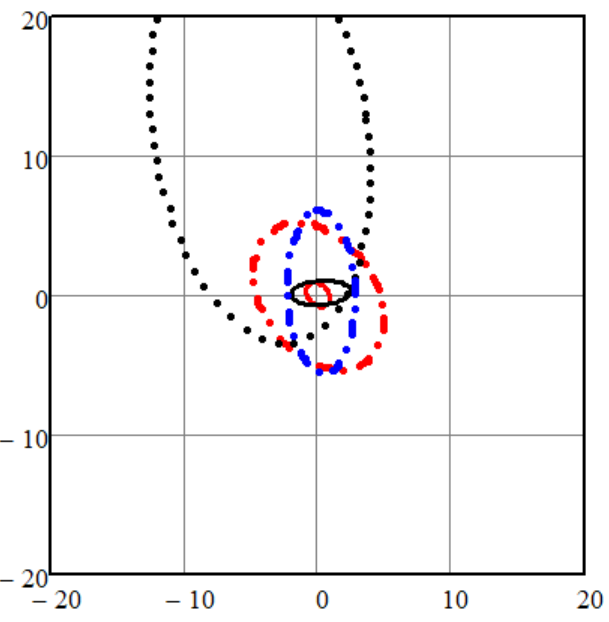}{0.295\textwidth}{(c)} 
          }
\caption{(a) Full (417\arcsec by 417\arcsec) image of the 8 component bulge model (listed in Table~\ref{tab:BNmodel}).
(b) Large (800\arcsec by 800\arcsec) region showing the ellipses with semi-major axis $R_e$ for the three Sersic components of the bulge model. 
(c) Central (40\arcsec by 40\arcsec) region showing the ellipses with semi-major axis $FWHM/2$ for Gaussians or $R_e$ for the five components of the nuclear region model. 
Panels (b) and (c) have different scales than panel (a) in order to show the components with their greatly different sizes.
  }
\end{figure*}

Because the components of the multicomponent models have different centers, axis ratios and position angles,
the total model is expected to have some asymmetry.
We find that to be the case, as illustrated in panel (a) of Figure~\ref{fig:asym} for the best-fit 8-component model
from Table~\ref{tab:BNmodel}. 
The centres of each component, in general, are offset from the position of the M31 nucleus \citep{2002AJ....124..294P}.  
The model components for the nucleus (compact nucleus plus nuclear bulge) have low offsets. 
From Table~\ref{tab:BNmodel} the 8-component model has nucleus component offsets from 
$\simeq$0.08\arcsec to 12.6\arcsec, with only the most extended component (the second Gaussian) having an offset exceeding 3.2\arcsec.
Higher offsets occur for the centres of the bulge model functions. Generally the three Sersic components 
for the bulge (see Table~\ref{tab:BNmodel}) have offsets of $\sim$42\arcsec, $\sim$1\arcsec and  $\sim$16\arcsec. 
The offsets as a fraction of the $R_e$ values are $\sim$0.1, $\sim$0.01 and  $\sim$0.08, respectively.

The  M31 bulge model (sum of all components, Figure~\ref{fig:asym} panel a) exhibits an asymmetry in the bulge light distribution, consistent with the fractional offsets ($\sim0.1$)  listed above for the centres of the model components. 
The model has deviation from perfect elliptical symmetry and shows a faint extension to the NNW of the centre of the bulge. 
We illustrate how the individual components contribute to the shape and asymmetry of the bulge:
panel (b) shows the three large bulge model Sersics by their ellipses with semi-major axes $R_e$, including orientation and offsets;
panel (c) shows the nuclear region model with the four-Gaussian plus Sersic components.
The asymmetry and NNW extension are also seen in the original observation (panel (a) of Figure~\ref{fig:UVITposn}): however in the data, the extra confusion caused by the clumpiness of the bulge make it more difficult to discern.
The large number of components (8 in the final model) are needed in order to produce a good fit to the asymmetry and NNW extension.
Thus to a large extent it is a phenomenological model for a light distribution which differs from 
an elliptical Sersic distribution.
The meaning of the different components used to fit the bulge is discussed further below in Section ~\ref{sec:interp}.

\section{Discussion}\label{sec:disc}

\subsection{The Choice of GALFIT for Image Fitting}\label{sec:reliability}

GALFIT's performance has been tested in fitting brightness distributions of galaxies. 
\citet{2008ApJ...683..644S} used a sample of 50,000 simulated galaxies to find that GALFIT returned the Sersic index correctly
for more than 90$\%$ of the cases. 
 \citet{2017ApJ...844L...6P} tested the recovery of galaxy sizes with GALFIT 
 and found only small offsets between for both star-forming and quiescent galaxies. 
 \citet{2013MNRAS.430..330H} tested GALFIT on simulated data and found recovery of parameters, including magnitude. 
Using simulating galaxies from the Sloan Digital Sky Survey, 
\citet{2013MNRAS.433.1344M} find that single Sersic function parameters are recovered with $\sigma_{mag}$ $\approx$ 0.025 mag and $\sigma_{radius}$ 	$\approx$ 5 percent. 
\citet{2007ApJS..172..615H,2008AIPC.1082..137H} 
used simulated galaxies to test the performance of GALFIT against GIM2D, another two-dimensional fitting algorithm. 
Their results suggest that GALFIT outperforms its counterpart in almost all of the tested categories. 
The performance tests conducted over the years on GALFIT have established it as a benchmark software for 
two-dimensional galaxy modelling. 

New fitting algorithms that are designed on neural network and deep learning computational principles frequently use GALFIT to test their software. 
For example, \citet{2018MNRAS.475..894T} use Deep Neural Networks to study galaxy morphology and show that their new algorithm produces results as good as GALFIT, but much faster.

\subsection{Comparison of Single Sersic Fits with Previous Model Fits of the M31 Bulge}\label{sec:compare}

The other UVIT filter images (F169M, F172M, N219M and N279N) for the 
the M31 bulge region (inner $\simeq$1.59 kpc by 1.59 kpc) were fit with single Sersic functions.
The small central nuclear region is omitted for these fits. 
The resulting single Sersic fit parameters, including the F148W fit, are given in Table~\ref{tab:1Sersic}. 
The Sersic indices range from $n=2.14\pm0.05$ at 279 nm to $n=2.51$ to 2.66 for the 148 to 219 nm data;
the effective radii are $R_e=0.46$ to 0.51 kpc; and the ellipticities are $\epsilon=1-q=0.25$ to 0.27.  

\begin{deluxetable*}{ccccccccccc}
\tablecaption{One Component Sersic Parameters for the Five UVIT Images \label{tab:1Sersic}}
\tablewidth{700pt} 
\tabletypesize{\scriptsize}
\tablehead{\colhead{Image} & \colhead{Component} & \colhead{Magnitude} & \colhead{$R_e$(\arcsec or kpc$^a$)} & \colhead{Sersic Index } & \colhead{Axis Ratio } & \colhead{Position Angle($^\circ$)}\\ }
\startdata F148W & Sersic & 11.78$\pm$0.02 & 127$\pm$3(0.483$\pm$0.011) & 2.52$\pm$0.02 & 0.730$\pm$0.001 & -48.3$\pm$0.1 \\
F169M & Sersic & 12.43$\pm$0.05 & 120$\pm$6(0.457$\pm$0.023) & 2.51$\pm$0.05 & 0.740$\pm$0.002 & -48.1$\pm$0.3 \\
F172M & Sersic & 13.30$\pm$ 0.06 & 134$\pm$8(0.510$\pm$0.030) & 2.66$\pm$0.07 & 0.750$\pm$0.002 & -48.3$\pm$0.3 \\
N219M & Sersic & 12.05$\pm$0.09 & 176$\pm$15(0.670$\pm$0.057) & 2.52$\pm$0.08 & 0.770$\pm$0.002 & -48.2$\pm$0.4 \\
N279N & Sersic & 11.04$\pm$0.06& 144$\pm$8(0.548$\pm$0.030) & 2.14$\pm$0.05 & 0.750$\pm$0.002 & -48.1$\pm$0.3 \\
 &  &  &  &  &  & \\
\enddata
\tablecomments{a: $R_e$ in kpc is given in round brackets.}
\end{deluxetable*}

A number of previous studies have studied the structural properties of M31, including the bulge.
\cite{2008MNRAS.389.1911S} fit the Spitzer 3.6 micron 1-D profile with a Sersic bulge plus and exponential disk, finding Sersic index $n=1.71\pm0.11$ and $R_e=1.93$ kpc. 
\cite{2011ApJ...739...20C} fit both minor and major axis profiles combining the infrared data (either Spitzer 3.6 micron or I band (0.9 micron) data from \citealt{2002AJ....124..310C}) with star count data added for the outer regions. 
They find (their Table 2) Sersic indices between $n=2.0$ and 2.6 for the major axis with $R_e=0.8$ to 1.2 kpc, with both ranges including uncertainties.
For the minor axis, they find Sersic indices between $n=1.7$ and 2.4 with $R_e=0.2$ to 0.8 kpc.
Combined minor and major axis fits yield Sersic index $n=2.18\pm0.06$ with $R_e=0.82\pm0.04$ kpc with
ellipticity $\epsilon=0.28\pm0.01$ for the Spitzer data, and  Sersic index $n=1.83\pm0.04$ with $R_e=0.74\pm0.02$ kpc and $\epsilon=0.28\pm0.01$ for the \citet{2002AJ....124..310C} data.
Thus wide ranges of Sersic index, $R_e$ and $\epsilon$ are found, depending on the data used and on the fitting method. 

\cite{2013ApJ...779..103D} analyze the I band data from \citet{2002AJ....124..310C} combined with star data (counts and kinematics) to fit structural parameters, luminosity functions and disk fractions (based on kinematics).
They find (their Table 2) a Sersic index $n=1.92\pm0.08$ and $R_e=0.78\pm0.03$ kpc for the bulge, with
ellipticity $\epsilon=0.28\pm0.01$.
This confirms these values for the I band, consistent with \cite{2011ApJ...739...20C}, which are different  (smaller $n$ and $R_e$ and larger $\epsilon$) than values derived using 3.6 micron data.

In summary, previous works 
in I-band and 3.6 micron yield single Sersic fits to the bulge with 
Sersic indices from 1.7 to 2.6, with most errors $\sim0.05$ to 0.2, and major axis $R_e$ from 0.7 to 1.9 kpc, with most errors $\sim0.01$ to 0.1. 
Fits including both major and minor axes find $\epsilon$=0.21 to 0.28, with error 0.01. 
The UVIT FUV and NUV results are not so different:
the UVIT single-Sersic indices are higher except for the N279N filter, the $R_e$ values are smaller and the $\epsilon$
values the same as the I and 3.6 micron band values. 
The differences are not surprising given the difference in waveband: cooler stars are more prominent in I and 3.6 micron bands, whereas hot stars are more prominent in the NUV and FUV bands. 
Hot star emission has a major contribution from Extreme Horizontal Branch stars, Post-AGB stars, Post-Early AGB stars, and AGB-manque stars \citep{2012ApJ...755..131R}, which are likely distributed differently than the cool stars.

\subsection{Interpretation of Multiple Component Fits}\label{sec:interp}

In modelling the UVIT F148W image of the M31 bulge, we carried out a series of tests. 
The dust lanes have a significant effect, so were masked (not included) in the fits (Section~\ref{sec:dustmask}).
Several bright stars affect the fits, so are included separately (Section~\ref{sec:background}). 
The nuclear region is a separate component, so when modelling the bulge only, the nuclear region is masked (Sections~\ref{sec:nucmask} and \ref{sec:bulgemodel}).
Then we fit the bulge with a single Sersic function and multiple Sersic functions.
We carried out fits including the nuclear region (Section~\ref{sec:nucmodel}), with model functions for both bulge and nuclear region, where the nuclear region includes the compact nucleus plus the nuclear bulge \citep{2002AJ....124..294P}.
The most relevant fits for comparison with previous work on models for the M31 bulge are the single component Sersic fits, listed in Table~\ref{tab:1Sersic}.

The purpose of the multicomponent fits to the bulge and nuclear region was to obtain a more accurate, although much more complex, functional description of the asymmetric surface brightness of the bulge and nuclear region. 
From Table~\ref{tab:BNmodel}, with 3 Sersic components for the bulge and 1 to 5 components for the nuclear region, the FUV (F148W) light distribution is seen to be complex and better described by multiple components than a single component.
These components are a purely phenomenological description of the F148W light distribution. 
They were derived from fitting the 2-dimensional light distribution only, without constraints from other data,
from simulations or from theory. 

The 3 Sersic components for the F148W emission from the bulge, with different models for the nuclear region, consistently have $R_e$ values of 0.33, 0.75 and 1.5 kpc and respective Sersic indices of 1.1, 1.5 and 0.3. 
The 3 Sersic functions have different axis ratios, different position angles and different centers (by up to
42\arcsec or 0.16 kpc for the Sersic function with $R_e$= 1.5 kpc) in order to fit the asymmetry in the light distribution of the bulge (see Figure~\ref{fig:asym} panels (a) and (b)).
For the bulge region, it is known to be complex from the work of \citep{2002AJ....124..294P} and references therein. 
The five component model (model 5 in Table~\ref{tab:BNmodel}) for the nuclear region is illustrated in panel (c) of Figure~\ref{fig:asym}. 
This fit is a phenomenological model for the complex FUV light distribution in the nuclear region.
The fit was carried out simultaneously with the 3 Sersic model for the bulge in order to separate, as best as possible, the nuclear region light from the bulge region light in the nuclear region.  

\textbf{
The M31 bulge light distributions in the other UVIT filters (F169M, F172M, N219M and N279N) were fit with multi-component models. The results of those fits are given in the Appendix and Table~\ref{tab:4compFits}.  With the lower signal-to-noise for those images, the number of components was 4, and the parameters of those components are considerably less well constrained. 
}

\subsection{Bulge Type}\label{sec:bulgetype}

\subsubsection{Discussion of bulge type}\label{sec:bulgedisc}

The two bulge categories of pseudobulges and of classical bulges are essential  in understanding galaxy formation history (KK04, \citealt{2008AJ....136..773F}). 
According to KK04, the Sersic index 
can be used to classify bulges, or at least provide evidence in favor of one of the types in the absence of kinematic information. 
A Sersic index value less than 2 is consistent with a pseudobulge, while classical bulges have a characteristic Sersic index greater than 2 (KK04, \citealt{2008AJ....136..773F,2016ASSL..418...41F,2008MNRAS.384..420G}). 
\citet{2016ApJ...821...90A} distinguish i) classical bulges by having a spheroidal shape, Sersic index $>2.5$ and $V_{max}/\sigma$ values that are less than for isotropic oblate rotators; and ii) pseudobulges by having the shape of a disk, Sersic index $<2.5$ and clear signs of rotation. 

\citet{1993MNRAS.265.1013C} demonstrated that ellipticals and galactic bulges are better fit if the exponent of the $r^{1/4}$ term in the de Vaucouleurs formula is a free parameter as $r^{1/n}$, equivalent to the Sersic function with index n. 
Further studies, such as \citet{1996ApJ...457L..73C}, strengthened the idea that “late-type” bulges have a Sersic index between 1 and 2. 
This led KK04 to include n between $\approx$ 1 and 2 as one of the diagnostic tests for galactic pseudo-bulges.  
\citet{2008AJ....136..773F} performed bulge-disk decompositions on a sample of spiral galaxies and likewise concluded that Sersic indices smaller than 2 are indicative of a pseudobulge, while greater than 2 are indicative of a classical bulge.
This is confirmed by the \citet{2008MNRAS.384..420G} study on the structural parameters of bulges, bars and disks. 
They found that bulges show significantly different parameters than ellipticals, with a greater dispersion for pseudobulges.
\citet{2010ApJ...716..942F} find value of $n < 2$ may be interpreted as evidence of secular evolution in the bulge, and suggest that their n value close to 2 suggests “some amount” of secular evolution. 
Nine S0--Sb galaxies which have two-component bulges (classical plus pseudobulge) are the subject of an observational study by \citet{2015MNRAS.446.4039E}.  
They find the disky pseudobulges consist of an exponential disk with either nuclear rings, nuclear bars or spiral arms and that the classical bulge components have Sersic indices in the range 0.9 to 2.2. 

Simulations of galaxy evolution provide a guide to the structural differences between pseudobulges and classical bulges.
\citet{2011ApJ...742...76G} present a cosmological N-body hydrodynamic simulation  
in which the evolutionary pathway of a close analog to the Milky Way disk galaxy is studied.  
Bulge formation was observed to have occurred through slow secular processes by the interaction of the various structural components of the galaxy, rather than fast and energetic merger events, which were absent in the simulation. 
Their simulation outcome of a “present-day” disk shape with a complex structure showing nuclear, bulge, disk and halo regions, was subsequently subjected to a GALFIT decomposition. 
The resulting bulge Sersic index was n = 1.4. 
Simulations for a major merger of two gas-rich pure disk galaxies \citep{2012MNRAS.424.1232K}, finds the resulting bulge has Sersic index $\sim0.3$- 0.9.
When the initial galaxies have gas haloes \citep{2016ApJ...821...90A}, the immediate outcome is a classical bulge, but a pseudobulge and a thin disk grow from gradually accreted gas, which can be considered secular evolution. 
The kinematics and range of stellar ages for the classical and pseudobulge are very different.
The classical bulge (stars born at time $t<1.4$ Gyr) has a Sersic index of $n\sim4$-6, the 4 remaining components (stars born at time intervals $1.4<t<1.8$ Gyr, $1.8<t<2.2$ Gyr, $2,2<t<9$ Gyr and $9<t<10$ Gyr) ) can be fit with a combination of 3 exponential disks. 

A second characteristic of pseudobulges as described by KK04, \citet{1996ApJ...457L..73C} and  \citet{2010ApJ...716..942F} is the presence of boxiness. Bulges showing box-shaped isophotes are well-established observationally and theoretically \citep{1961ApJ...133..355S,1974IAUS...58..335D}. 
Boxy bulges are observed in about one fifth of edge-on spiral galaxies \citep{2016ASSL..418...77L,2000A&AS..145..405L,1987A&AS...70..465D}. 
KK04 conclude that the detection of boxy bulge isophotes is sufficient for the identification of a pseudobulge. 
This stems from the connection boxy bulges seem to have with galactic bars: they appear to be created by the interaction of the bar with the disk as secular evolutionary mechanisms slowly construct the boxy bulge out of disk material. 

\subsubsection{M31's Bulge}\label{sec:boxy}

For M31, \citet{2011ApJ...739...20C} 
use one-dimensional decomposition methods to model the bulge,  and obtain a Sersic index n = 2.2$\pm$3. 
They measured the bulge scale-length to disk scale-length ratio of 0.2  which agrees with the value of 0.2 
which is obtained from secular evolution models \citep{1996ApJ...457L..73C}. 
\citet{2011ApJ...739...20C} note that their Sersic index is close enough to 2 to be consistent with some amount of secular evolution in the bulge, according to \citet{1996ApJ...457L..73C}, \citet{2010ApJ...716..942F} and  KK04. 
The study here shows that a single Sersic fit to the five UVIT images (Table~\ref{tab:1Sersic}) yields 
indices between $\sim2.1$ and  $\sim2.6$. 
Here we used two-dimensional image fitting with GALFIT, which should better model the light distribution than one-dimensional fits (see Section~\ref{sec:reliability} above).
We obtain a M31 bulge Sersic index of 2.14 for the 279nm UVIT image, consistent with earlier results, and 
indices of $\simeq$2.5 for the four shorter wavelength UVIT images (148 to 219 nm). 
The fact that the Sersic index is $\simeq$2 to 2.5 and neither clearly $<2$ or clearly$>2.5$ indicates there are probably contributions from a classical bulge and a pseudobulge in M31. 
The four shorter wavelength images yield a larger Sersic index, possibly indicated a smaller contribution from a pseudobulge relative to the classical bulge at shorter wavelengths.

Several studies indicate that the M31 bulge is intrinsically an oblate spheroid with an axis ratio of $\sim$0.8 \citep{2006MNRAS.366..996G,1983ApJ...266..562K,2003ApJ...588..311W} while others indicate a triaxial nature \citep{1956StoAn..19....2L,1977ApJ...213..368S,1989AJ.....97.1614K,1994ApJ...426L..31S,2001A&A...371..476B,2002MNRAS.336..477B}. 
For both cases, two-dimensional fitting of the image is superior to one-dimensional fitting.

\begin{figure*}[htbp]
    \centering
    \epsscale{0.85}
 \plotone{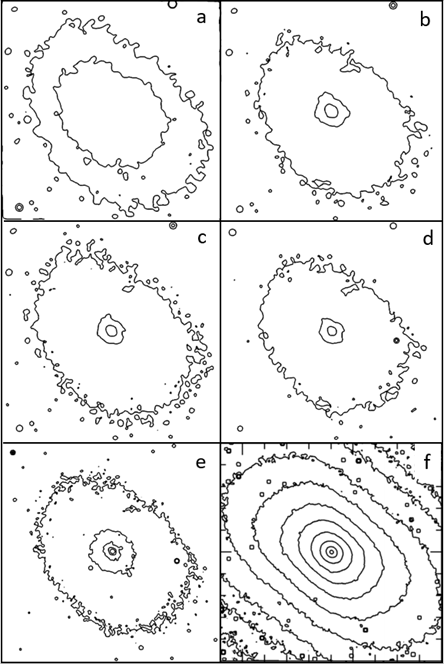}
   \figurenum{5}
    \label{fig:contour}
    \caption{Contour lines for the UVIT images of the bulge of M31 in the (a) F148W, (b) F169M, (c) F172M, (d) N219M and (e) N279N filters.  On the same scale, the contours from   
    \citet{2007ApJ...658L..91B} are shown for comparison in (f). The boxy nature of the contours is evident at all wavebands. 
    }
\end{figure*}

\begin{deluxetable*}{cccccc}
\tablecaption{Generalized Ellipse$^a$ (C$_0$) and Fourier Mode 4$^b$ Measures of Boxiness\label{tab:boxy}}
\tablewidth{700pt} 
\tabletypesize{\scriptsize}
\tablehead{\colhead{Image} &  C$_0$ &  m=4 Fourier mode&   \\
  & & Amplitude & Phase Angle($^\circ$) \\ }
\startdata  
F148W & 0.193$\pm$0.007 & 0.0150$\pm$0.0006 & -1.5$\pm$0.6 \\
F169M & 0.159$\pm$0.019 & 0.0124$\pm$0.0012 & 1.2$\pm$1.7 \\
F172M & 0.187$\pm$0.023 & 0.0144$\pm$0.0014 & 0.6$\pm$1.7 \\
N219M & 0.098$\pm$0.000 & 0.0085$\pm$0.0012 & 9.4$\pm$2.9 \\
N279N & 0.156$\pm$0.018 & 0.0121$\pm$0.0011 & 4.4$\pm$1.6 \\
\enddata
\tablecomments{a: Diskiness has C$_0<0$; boxiness has C$_0>0$. b: The amplitude of the m=4 Fourier mode is $>0$ for boxy ellipses and $<0$ for disky ellipses.}
\end{deluxetable*}

A near-infrared mapping study of M31 conducted by \citet{2007ApJ...658L..91B} revealed the boxy morphology of the bulge with high S/N and with little interference from dust extinction. 
The UVIT FUV and NUV images of the bulge of M31 show a boxy morphology. 
Figure~\ref{fig:contour} shows contour lines from the five UVIT images (panels a through e). 
Panel f shows the contour lines from \citet{2007ApJ...658L..91B} which illustrate similar boxiness to  that seen in the N279N image in panel e. 
The other UVIT filter images exhibit boxiness, although less strongly than that seen for the N279N image.

Most of the boxy or peanut shaped features which have been reported in galaxies have been detected in infrared bands \citep{2016ASSL..418...77L,2010ApJ...715L.176K}, such as Spitzer 3.6 micron images.
In the UV there have been no previous reports of bulge boxiness. 
However, all five of the UVIT filter images of the bulge of M31 show signs of a boxiness.
Considering KK04’s third criterion, the bulge boxiness indicates a pseudobulge interpretation for M31's bulge. 

There are two ways to measure diskiness or boxiness from fitting images using GALFIT.
The first method is the use of generalized ellipses instead of ellipses, as follows.
The radial functions are elliptical in shape, accomplished by making the radial coordinate $r$ constant on ellipses, defined by:
\begin{equation}
r(x,y) = \left( |x-x_{c}|^{2} + | \frac{y-y_{c}}{q}|^{2} \right) ^{\frac{1}{2}}. 
\end{equation}
To allow the ellipse to have arbitrary orientation, $(x,y)$ are rotated coordinates.
To change this into generalized boxy or peanut shaped ellipses, the power is changed from
2 to ${C_{o}+2}$ as follows \citep{2016ApJ...821...90A}:
\begin{equation}
r(x,y) = \left( |x-x_{c}|^{C_{o}+2} + | \frac{y-y_{c}}{q}|^{C_{o}+2} \right) ^{\frac{1}{{C_{o}+2}}}.
\end{equation}
$C_{o}$ is the parameter which controls the diskiness or boxiness of the isophotes (lines of constant $r(x,y)$). 
For $C_{o} = 0$ the isophotes are perfect ellipses. 
For $C_{o} < 0 $ the isophotes becomes more disk-like or disky as $C_{o}$ becomes more negative.
For $C_{o} > 0 $ the isophotes becomes more boxy as $C_{o}$ becomes more positive. 
Table~\ref{tab:boxy} shows results of fitting generalized ellipses for the five UVIT images. 
All values of $C_{o}$ are $>0$, showing that the M31 bulge has a boxy shape. 

The second method is to fit radial functions modified by azimuthal functions in GALFIT, as described in \citet{2010AJ....139.2097P}.  
The $m=4$ fourier mode describes whether the bulge is boxy or disky.
Table~\ref{tab:boxy} shows the results of fitting the bulge including the m=4 fourier. 
For all five UVIT images the boxiness of the bulge is significant (amplitude $>0$ at 7 to 25$\sigma$).
The Fourier mode and $C_{o}$ tests of boxiness are consistent for each image: boxiness is strongest for the F148W image, 
less strong for the F172M image, then the F169M and N279N images, and weakest but still significant for the N219M image.

\subsection{The nuclear spiral and dust absorption features}\label{sec:spiral}

The nuclear spiral is a noticeable feature of the central 4 arcmin of the bulge of M31 at specific wavelengths. 
 It is observable in optical emission lines SII, OIII \citep{1985ApJ...290..136J} and H$_\alpha$ \citep{2009MNRAS.397..148L}, in radio \citep{1985A&A...150L...1W},
in infrared emission \citep{2009MNRAS.397..148L} and as extinction features found by analysis of broadband optical filters \citep{2016MNRAS.459.2262D}.  
The nuclear spiral increases in radius in a clockwise direction. 
It is not clear from optical images which direction the outer main spiral arms increase in radius (e.g. using the Digitized Sky Survey images of M31). 
Doppler shifts \citep{1970ApJ...159..379R} clearly indicate that the NE of M31 is moving away from the observer and the SW moving toward the observer.
\cite{2017A&A...600A..34T} use deprojected Herschel 250 micron and GALEX NUV images to visualize the spiral arms structure of M31. Their work indicates that the arms increase in radius in a clockwise direction, similar to the nuclear spiral, but with a smaller increase in radius with angle. 

In ultraviolet, the bulge is dominated by smooth emission (e.g. Fig. 1). 
In order to see the fine structure of the central region of the bulge in ultraviolet one has to subtract the large scale distribution of the light.
We subtracted our best-fit bulge model to obtain the residual image of the central part of M31. 
The sequence of images in Figure~\ref{fig:resid} shows that as the bulge model improves, the dust lanes in the centre of the bulge become more visible.

\begin{figure*}[htbp]
    \centering
    \epsscale{1.2}
 \plotone{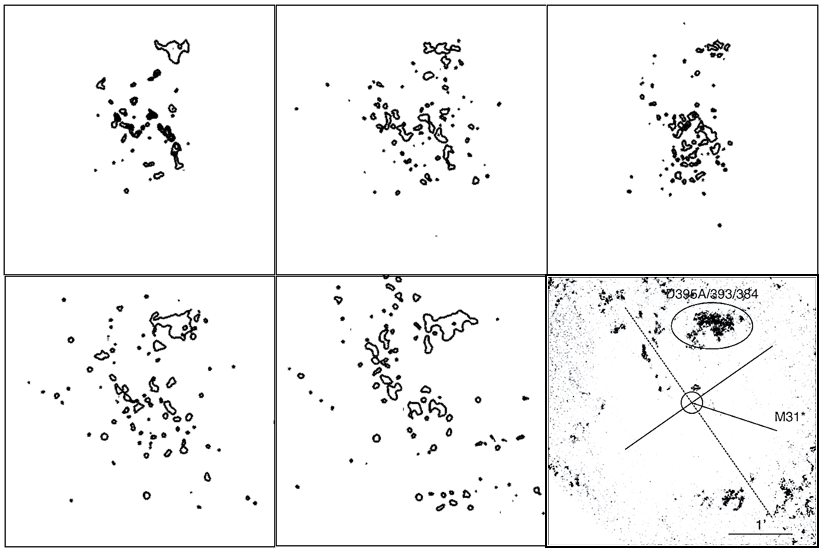}
    \figurenum{6}
    \label{fig:spiral}
    \caption{The nuclear spiral dust lane features: negative contours for the UVIT residual images (after subtraction of the best-fit bulge model) are shown in the first 5 panels: (a) F148W, (b) F169M, (c) F172M (left to right in top row), 
     (d) N219M, (e) N279N (left and centre in bottom row); 
    the extinction map from \citet{2016MNRAS.459.2262D} in greyscale is shown in the bottom right panel (f). The ellipse in panel (f) marks a foreground dust cloud.}
\end{figure*}

\begin{figure*}[htbp]
    \centering
    \epsscale{1.2}
 \plotone{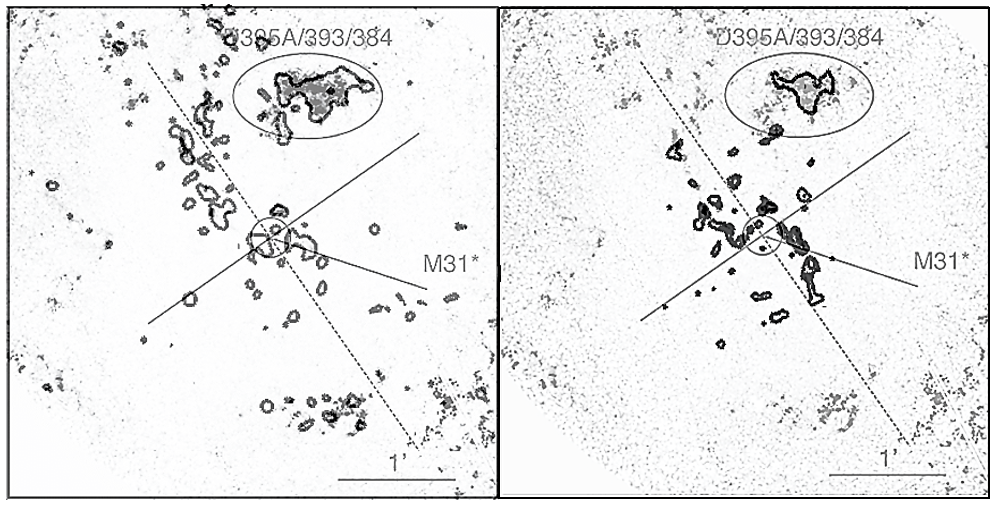}
    \figurenum{7}
    \label{fig:overlay}
    \caption{Overlays of nuclear spiral dust lane features from UVIT on the greyscale extinction map of \citet{2016MNRAS.459.2262D}: N279N residual image negative contours (left panel) and  F148W residual image negative contours (right panel).  
    }
\end{figure*}

The nuclear spiral is $\sim$4 arcmin diameter, thus we show in Figure~\ref{fig:spiral} the contour lines of residual images (size 250\arcsec by 250\arcsec) around the centre of M31 for the five UVIT filters. 
The contour lines are at negative values in the residual images to emphasize the dust absorption features.
The sixth panel of Figure~\ref{fig:spiral} is the extinction map from \citet{2016MNRAS.459.2262D} for comparison.
Figure~\ref{fig:overlay} shows the overlay of the inner 250”x250” of the UVIT residual image 
 contour lines  and the extinction map of \citet{2016MNRAS.459.2262D}.  
The nuclear spiral is clearly visible. 
Radial velocity studies show that the nuclear spiral is offset radially from the M31 nucleus by less than 50 pc, placing the spiral near the centre of bulge \citep{2016MNRAS.459.2262D}. 
The large clump (marked with an ellipse) north of the centre is a dust cloud in front of the bulge along our line of sight \citep{2000MNRAS.312L..29M}.
The close match between the dust lanes detected in the UVIT images and the extinction map of  \citet{2016MNRAS.459.2262D}  serves as a verification of our modelling procedure and the best-fit bulge models obtained. 

\section{Conclusion}\label{sec:conc}

The nuclear region of M31 is dynamically and photometrically distinct from the bulge and the large scale galactic disk \citep{1993AJ....105.1793K,1993AJ....106.1436L}.
Here  we construct models of the luminosity distribution of the bulge at NUV and FUV wavelengths using UVIT observations with 1\arcsec spatial resolution.
The initial fits omit the nuclear region, then we include the nuclear region. 
The best-fit multicomponent model for the F148W image, which has highest sensitivity of the five UVIT images, has eight components 
with three components for the bulge and five for the nuclear region (Table~\ref{tab:BNmodel}). 
These components have different centers, sizes and ellipticities, and are required to fit the asymmetric shape of M31's bulge. 
The 8 component model very well describes the smooth part of the light-distribution so that the residual image can be used to study the sharper features such as dust absorption lanes. 

One component models of the light distribution for M31's bulge region are useful to compare to previous studies at other wavelengths. 
The Sersic indices for the five UVIT FUV and NUV bands are: $\simeq$2.5 for 148, 169, 172 and 219 nm and
$\simeq$2.1 for 279 nm. 
The 279 nm value is consistent with Sersic indices found for M31 previously.
The $R_e$ values are $\simeq$0.46 to 0.67 kpc, which are slightly smaller than values found from I band and 3.6 micron images.
The observed images and the model images show strong boxiness, which is confirmed by fitting using boxiness parameters in GALFIT. 
The Sersic indices are intermediate between typical values for a classical bulge $>2.5$ and typical values for a pseudobulge ($\sim1$ to 2), which indicates that M31's bulge likely has both classical and pseudobulge components.
The boxiness also argues for a significant pseudobulge component. 

Residual images for the five UVIT images were generated by subtracting the best-fit model for each filter from the observed   brightness distribution. 
The residual images were used to map the dust lanes associated with the nuclear spiral which would otherwise be hidden by the bright starlight in the bulge region. 
The FUV and NUV dust lanes match closely the dust lanes seen in optical extinction \citep{2016MNRAS.459.2262D}.

There is substantial agreement between our results and the conclusions of other studies on the nature of the bulge of M31 \citep{2018MNRAS.481.3210B,2018A&A...618A.156S} and on the location of the nuclear spiral dust lanes \citep{2016MNRAS.459.2262D,2009MNRAS.397..148L}. 
This serves as a verification of our model fitting procedure and model results.
In future studies, we plan to take advantage of the 1\arcsec resolution of the UVIT images to study the complex structure of the other 
components of M31, such as the disk, spiral arms and halo. 

\begin{acknowledgments}
This work was supported by a grant from the Canadian Space Agency. The authors thank the reviewer for making a number of suggestions to improve this manuscript.
\end{acknowledgments}

\appendix

\section{Models for the M31 bulge for UVIT F169M, F172M, N219N, and N279N data}\label{sec:otherfilter}

Similar to the F148W image fitting, the four UVIT images (F169M, F172M, N219M and N279N) were fit with 2-dimensional
 brightness profiles with inclusion of the dust mask (exclusion of the dust lanes) and with fits for the brightest stars in the image (similar to Table~\ref{tab:stars} for the F148W image).
 One-component fits were carried out including the nucleus mask (i.e., excluding the nuclear region) with a Sersic function in order to be able to compare to the one component Sersic function fit for the F148W image.

Table~\ref{tab:1Sersic} lists the parameters for the one component Sersic model fits for all five images,
including F148W.  
The typical Sersic index for the one-Sersic fits (Table~\ref{tab:1Sersic}) is $\simeq$2.5 for the four shortest wavelength filters, but close to 2.1 for the longest wavelength N279N filter.  
All filters yield the same position angle and nearly the same axis ratio for the bulge.
Possible reasons that  the N279N fit is somewhat of an outlier in Sersic index is that it is sensitive to different stars in a significant way than the shorter wavelength NUV and FUV filters. E.g. as noted in Section~\ref{sec:compare}, the short wavelength filters are sensitive to hot stars (such as Extreme Horizontal Branch stars) whereas the N279N is more sensitive to intermediate temperature stars.  

Like the F148W image, the other four UVIT images are significantly better fit by including additional model components.
For this set of fits the nuclear region is included (we omit the nuclear mask).
Rather than carrying out the full procedure outlined in Figure~\ref{fig:flow}, we only tested Sersic functions
rather than all five functions listed in Section~\ref{sec:profiles}.
This was done for the following reasons: the process of fitting for the F148W image took several months, because of the large number of models tested; the goal is to have similar functions for all five images in order to fairly compare the parameters; and the F169M, F172M, N219M and N279N images are of lower signal-to-noise.
The simplest multiple-Sersic model which gave a good fit for the F169M, F172M, N219M and N279N images was a three Sersic function model 
for the bulge plus a single Sersic function for the nuclear region, called the 3+1 Sersic model here.
Because the F169M, F172M, N219M and N279N images were lower signal-to-noise than the F148W image (by factor of $\sim$2 to 3, see Table~\ref{tab:obs}), models more complex than four Sersic components were not justified. I.e. the statistical improvement in the fits was not significant.
The results of the 3+1 Sersic model fits are shown in Table~\ref{tab:4compFits}.

\begin{deluxetable*}{ccccccccccc}
\tablecaption{3+1 Sersic Parameters for the F169M, F172M, N219M and N279N Images$^a$ \label{tab:4compFits}}
\tablewidth{700pt} 
\tabletypesize{\scriptsize}
    \tablehead{\colhead{Image} & \colhead{Component} & \colhead{Magnitude} & \colhead{$R_e$(\arcsec or kpc$^b$)} & \colhead{Sersic Index} & \colhead{Axis Ratio } & \colhead{Position Angle($^\circ$)} & \\ }
\startdata  
F169M & Bulge &   &   &   &   &   \\
  & Sersic & 13.4$\pm$0.3 & 480$\pm$120(1.83$\pm$0.46) & 0.87$\pm$0.08 & 0.410$\pm$0.003 & -30.6$\pm$0.6 \\
    & Sersic & 13.47$\pm$0.06 &119$\pm$5(0.452$\pm$0.019) & 1.84$\pm$0.03 & 0.780$\pm$0.002 & -57.4$\pm$0.4 \\
    & Sersic & 14.94$\pm$0.45 & 420$\pm$170(1.60$\pm$0.65) & 0.36$\pm$0.02 & 0.200$\pm$0.003 & 75.9$\pm$0.7 \\
  & Nucleus &   &   &   &   &   \\
  & Sersic & 18.34$\pm$0.02 &3.1$\pm$0.1(0.0120$\pm$.0004) & 2.44$\pm$0.12 & 0.87$\pm$0.03 & -49.3$\pm$8.6 \\
  &   &   &   &   &   &   \\
F172M & Bulge &   &   &   &   &   \\
  & Sersic & 14$\pm$7 &2530$\pm$170(9.63$\pm$0.66) & 0.42$\pm$0.03 & 0.04$\pm$0.24 & -39.8$\pm$0.6 \\
    & Sersic & 14.2$\pm$0.4 & 265$\pm$33(1.00$\pm$0.13) & 1.70$\pm$0.12 & 1.00$\pm$0.01 & n/a \\
  & Sersic & 14.2$\pm$0.6 & 250$\pm$120(0.95$\pm$0.46) & 3.46$\pm$0.36 & 0.750$\pm$0.003 & -54.4$\pm$0.7 \\
  & Nucleus &   &   &   &   &   \\
  & Sersic & 20.61$\pm$0.03 & 0.99$\pm$0.07(.0038$\pm$0.0003) & 1.16$\pm$0.28 & 0.88$\pm$0.07 & 67$\pm$24 \\
  &   &   &   &   &   &   \\
N219M & Bulge &   &   &   &   &   \\
  & Sersic & 10.34$\pm$2.9 & 1138$\pm$12(4.331$\pm$0.046) & 0.81$\pm$0.14 & 0.73$\pm$0.02 & -20$\pm$5 \\ 
   & Sersic & 12.50$\pm$1.0 & 660$\pm$370(2.5$\pm$1.4) & 1.02$\pm$0.17 & 0.46$\pm$0.01 & -46.3$\pm$0.6 \\
  & Sersic & 13.91$\pm$0.05 & 95$\pm$3(0.362$\pm$0.011) & 1.42$\pm$0.02 & 0.860$\pm$0.003 & -56$\pm$1 \\
  & Nucleus &   &   &   &   &   \\
  & Sersic & 18.36$\pm$0.07 & 5.3$\pm$0.7(0.020$\pm$0.003) & 3.57$\pm$0.34 & 0.78$\pm$0.04 & -49$\pm$6 \\
  &   &   &   &   &   &   \\
N279N & Bulge &   &   &   &   &   \\
  & Sersic & 11.86$\pm$0.18 & 242$\pm$2(0.921$\pm$0.008) & 1.82$\pm$0.01 & 0.730$\pm$0.004 & -31.7$\pm$0.6 \\
    & Sersic & 11.97$\pm$0.63 & 500$\pm$160(1.90$\pm$0.61) & 0.93$\pm$0.10 & 0.70$\pm$0.078 & -55$0	\pm$8 \\
  & Sersic & 12.72$\pm$0.26 & 295$\pm$41(1.12$\pm$0.16) & 1.62$\pm$0.07 & 0.530$\pm$0.005 & -73.8$\pm$0.7 \\
  & Nucleus &   &   &   &   &   \\
  & Sersic & 16.58$\pm$0.54 & 40$\pm$34(0.15$\pm$0.13) & 6.8$\pm$1.6 & 0.67$\pm$0.04 & 49$\pm$5 \\
\enddata
\tablecomments{a: The parameters for the 3+1 Sersic model for the F148W image are given by the 4 Component model in Table~\ref{tab:BNmodel}. b: $R_e$ in kpc is given in round brackets.}
\end{deluxetable*}

Comparison of the fits between different filters for the 3+1 Sersic models shows a significant variation in Sersic indices. 
As discussed in Section~\ref{sec:interp} above, the interpretation of the multiple-component fits is not clear: it is basically a phenomenological model for the asymmetric light distribution. 
The 3+1 Sersic models have components with a range of indices (excluding the nucleus Sersic index) 
from $\sim$0.4 to 3.5, with all indices but one (the index for the third component of the F172M image) less than 2.

\end{document}